\documentclass{aa}
\usepackage{graphicx}
\usepackage[breaklinks, colorlinks, citecolor=blue, urlcolor = blue]{hyperref}
\usepackage{multirow}
\usepackage{adjustbox}

\usepackage[varg]{txfonts}

\bibpunct{(}{)}{;}{a}{}{,} 

\newcommand \sw{{\it Swift}}
\newcommand \fe{{\it Fermi}}
\newcommand \eb{{$E_{\rm break}$}}
\newcommand \ep{{$E_{\rm peak}$}}
\newcommand{\order}[1]{} 
\begin{document}

\title{Insights into the physics of Gamma-Ray Bursts \\ from the high-energy extension of their prompt emission spectra}

\titlerunning{high-energy extension of GRB prompt emission}
\author {M.~E. Ravasio\inst{\ref{inst1},\ref{inst2}}
 \and G. Ghirlanda\inst{\ref{inst2}}
\and G. Ghisellini\inst{\ref{inst2}}
}

\institute{Department of Astrophysics/IMAPP, Radboud University, P.O. Box 9010, 6500 GL, Nijmegen, The Netherlands \\
\email{\href{mariaedvige.ravasio@ru.nl}{mariaedvige.ravasio@ru.nl}}\label{inst1}
\and
INAF–Astronomical Observatory of Brera, via E. Bianchi 46, I-23807 Merate, Italy \label{inst2}
}  
\date{Received 04/12/2023 / Accepted: 23/02/2024}

\abstract
{
The study of the high-energy part (MeV-GeV) of the spectrum of gamma-ray bursts (GRBs) can play a crucial role in investigating the physics of prompt emission,  but it is often hampered by low statistics and the paucity of GeV observations. 
In this work, we analyze the prompt emission spectra of the 22 brightest GRBs which have been simultaneously observed by Fermi/GBM and Fermi/LAT, spanning six orders of magnitude in energy. 
The high--energy photon spectra can be modeled with a power-law $N(E)\propto E^{-\beta}$ possibly featuring an exponential cutoff. 
We find that, with the inclusion of the LAT data, the spectral index $\beta$ is softer than what is typically inferred from the analysis of Fermi/GBM data alone. Under the assumption that the emission is synchrotron, we derived a median value of the index $p\sim 2.79$ of the power-law energy distribution of accelerated particles ($N(\gamma) \propto  \gamma^{-p}$). 
In nine out of 22 GRB spectra, we find a significant presence of an exponential cutoff at high energy, ranging between 14 and 298 MeV.
By interpreting the observed cutoff as a sign of pair-production opacity, we estimate the jet bulk Lorentz factor $\Gamma$, finding values in the range 130–330.
These values are consistent with those inferred from the afterglow light curve onset time.
Finally, by combining the information from the high-energy prompt emission spectrum with the afterglow light curve, we exploited a promising method to derive the distance $R$ from the central engine where the prompt emission occurs. 
The distances ($R > 10^{13-15}$ cm) inferred for the only two GRBs in our sample that are suitable for the application of this method, which have only lower limits on their cutoff energies, suggest large emitting regions, although they are still compatible with the standard model. Larger samples of GRBs with measured cutoff energies and afterglow deceleration time will allow for more informative values to be derived.
These results highlight the importance of including high-energy data, when available, in the study of prompt spectra and their role in addressing the current challenges of the GRB standard model. 
}


\maketitle
\nolinenumbers

\section{Introduction}
Despite over 50 years of observations, the physical mechanism responsible for the prompt emission of gamma-ray bursts (GRBs) remains poorly understood, hindering our understanding of the Universe's most energetic and enigmatic phenomena. The fast variability (approximately tens or hundreds of milliseconds) of the prompt emission light curve, pointing toward a rather compact region, and the high luminosity of $\gamma$-ray photons of the prompt spectrum were the first pieces of observational evidence supporting the relativistic nature of GRBs, as their combination makes the compactness (defined as the total luminosity L emitted over the size R of the emitting region) too high to be compatible with observations \citep[e.g.,][]{Baring1997,Piran1999,lithwick01}. A relativistic motion, with $\Gamma\gtrsim 100$ for typical GRBs, is needed to overcome the compactness problem and it implies that we observed blue-shifted photons that are significantly softer in the co-moving rest frame.
The study of the prompt emission spectra offers a powerful tool to investigate the origin of the radiative mechanism, taking the relativistic motion into account. 
Thanks to the analysis of large samples of GRBs detected by the Burst And Transient Source Experiment (BATSE,  $\sim$\,20\,keV -- 2\,MeV) on board the Compton Gamma Ray Observatory (CGRO), the Gamma-ray Burst Monitor (GBM, $\sim$\,8\,keV -- 40\,MeV) on board \fe\ , the Burst Alert Telescope (BAT, 15 -- 150 keV), and the X--ray Telescope (XRT, $\sim$\,0.2\, -- 10\, keV) on board \sw, the typical observed GRB prompt spectrum appears nonthermal and it is often described  by two power-law functions, with indices $\alpha\,\sim\,-1$ and $\beta\,\sim \,-2.5$ at low and high energies, respectively, and they are smoothly connected at a characteristic (observer-frame) peak energy, \ep\,$\sim$\,300 -- 500\,keV, in the $\nu F_{\nu}$ representation
\citep[e.g.,][]{Preece2000,Kaneko2006,Nava2011,Goldstein12,Gruber2014}.

The origin of this powerful nonthermal emission, however, has been subject to debate for many years and is still uncertain.
A combination of a nonthermal component and a thermal one (also multiple thermal components in a few cases) has been adopted in some studies to model the observed spectra \citep[e.g.,][]{Ryde2010,Guiriec2011}. However, in most cases, the thermal component is subdominant and leaves the dominant nonthermal component unexplained.  
The interpretation of the prompt emission as being due to synchrotron faced for many years the inconsistency of the observed spectral shape with model predictions (see e.g., \citealt{Ghisellini2000}, but see the case of the Poynting-flux-dominated jet e.g., in \citealt{Zhang2009} and with a decaying magnetic field e.g., in \citealt{Uhm2014}). Only recently has the synchrotron interpretation been reconsidered thanks to the discovery of an additional spectral break at energies $\sim$2-200 keV in long \sw\ \citep{gor2017a,gor2018} and \fe\ GRBs \citep{Ravasio2018,Ravasio2019,Toffano2021} and the merger-driven GRB 211211A \citep{Gompertz2023}.
While these recent studies support synchrotron as the radiative mechanism of GRBs' prompt emission, the implied set of physical parameters opens new challenges. Interpreting the low-energy spectral break as the synchrotron cooling frequency, the implied magnetic field in the emitting region is small ($B^{\prime} \sim$ 10 G), requiring either large emitting regions (R$\sim 10^{16-17}$ cm) or large bulk Lorentz factors ($\Gamma \sim 10^3$), so as to avoid synchrotron self-Compton dominance and match the fast variability timescales (see also \citealt{Ghisellini2020,Florou2021}).
In this context, the high-energy part of the prompt emission spectrum can play a crucial role in addressing these challenges, giving access to fundamental physical quantities of the emitting region and of the acceleration mechanism.

However, the combination of poor statistics of the prompt spectrum at high energies (steep power law with a typical $\beta \sim -2.5$ of the Band function) and the low effective area of the available detectors makes it difficult to characterize precisely the spectrum above the peak energy. 
Besides the first few detections of radiation above several tens of MeV provided by the CGRO/EGRET \citep{Schneid1992,Schneid1995} 
and AGILE/GRID \citep{Giuliani2008,Moretti2009,Longo2009} 
instruments, most of our present knowledge on GRB high--energy emission comes from the Large Area Telescope (LAT, 100 MeV -- 300 GeV) on board \fe. 
Its larger field of view and effective area with respect to its predecessors allow for 14 GRBs to be detected per year on average,  corresponding to 
$\sim$ 12\% of the GBM-detected bursts (see \citealt{Nava2018} for a recent review on the high--energy emission of GRBs).
Usually, the LAT data reveal the presence of a high-energy emission
component, which rises with a small delay (approximately a few seconds) after the prompt emission, it lasts longer, and it is less variable. 
This emission is typically interpreted as afterglow, originating from shocks between the fireball and the 
interstellar medium (i.e., external shocks, see e.g., \citealt{Ghirlanda2010,Ackermann2010}). 
Instead, the early emission detected by LAT, which is simultaneous with the prompt phase and characterized by variability, can be interpreted as of an internal origin
\citep[e.g.,][]{Maxham2011,Zhang2011}, 
offering the exquisite opportunity to 
investigate the high--energy extension of the prompt emission spectrum. 
In particular, the inclusion of the LAT Low Energy (LLE, 30--100 MeV)  is crucial in bridging the gap between the GBM and the LAT data (>100 MeV), allowing a continuous characterization of the prompt emission spectrum.

When combining GBM and LAT data, the high--energy part of the prompt spectrum has been found to lie on the extrapolation of the keV--MeV (GBM) one, thus consisting of a single power law up to approximately GeV energies, although with a softer index than the one inferred from fitting the GBM data alone (e.g., \citealt{Ackermann2012,Abdo2009_080916C,Axelsson2012}). 
Assuming that the spectrum of accelerated electrons is a power law and that the cooling is dominated by the synchrotron emission, the photon index of the power law describing the spectrum above the peak energy allows one to constrain the spectral index $p$ of the electrons' energy distribution, providing fundamental insights into the acceleration mechanism \citep{Baring2009,Shen2006}.
On the other hand, it was also shown that the spectrum of a select number of GRBs requires a high--energy break between the GBM and LAT energy bands, revealing the presence of an exponential spectral cutoff from a few tens to hundreds of MeV \citep{Ackermann2012, Tang2015, Vianello2018}.  
The exponential drop in the spectrum is an important diagnostic tool as it could be interpreted as a sign of pair-production opacity, allowing one to estimate the bulk Lorentz factor $\Gamma$ of the jet \citep[e.g.,][]{lithwick01} and its compactness, or, alternatively, it could be linked to the maximum energy of accelerated particles. 
The knowledge of bulk Lorentz factor $\Gamma$ is crucial, as it allows one to infer the intrinsic properties of the emitting region, shedding light on many interesting - but still unknown - physical quantities (e.g., the location of the emitting region, the ejecta mass, the typical frequencies of the emitted radiation) that are fundamental to discriminate among different
theoretical scenarios. However, it is particularly difficult to obtain valuable constraints on $\Gamma$ from observations. So far, most of the measurements and lower limits currently available on $\Gamma$ have been derived from the detection of (or upper limit on) the peak time of the afterglow emission (see e.g., \citealt{Ghirlanda2018}).
The detection of an exponential cutoff in the prompt emission spectrum instead provides an alternative method to derive the bulk Lorentz factor $\Gamma$ of the jet. 
Moreover, the compactness argument is independent of the radiation mechanism producing the observed gamma-ray emission, giving constraints on $\Gamma$ and related emission parameters that have to be explained by any proposed prompt emission mechanism.

Therefore, from the study of the extension to high energies of GRB prompt emission spectra, finding either the smooth
continuation of the spectrum above the peak energy or the presence of a cutoff, 
we can gain valuable information regarding the physics underlying the GRB emission.
The combination of GBM and LAT data offers a powerful tool to investigate the GRB prompt emission spectra, providing the exquisite opportunity to characterize its shape over a wide energy range, spanning about 6 decades of energy.\\

In this work, we present the results of the joint analysis of GBM and LAT data of a sample of bright GRBs whose prompt emission has been detected up to GeV energies.
We describe the selection criteria and the analysis method in Section~\ref{sect:sample} and \ref{sect:methods}, respectively. The results of the spectral analysis are provided in Section~\ref{sect:results} and the interpretation of them in the context of the GRB standard model is given in Section~\ref{sect:interp}. In Section~\ref{sect:discussion}, we discuss the implication of our results, and how to exploit them to derive the emitting region distance from the central engine. We summarize the main findings in Section~\ref{sect:conclusions}.

\section{The sample}\label{sect:sample}

The second catalog of LAT-detected GRBs contains 186 events \citep{Ajello2019}, revealed during the first 10 years of activity of \fe. From the LAT online catalog\footnote{\url{https://heasarc.gsfc.nasa.gov/W3Browse/all/fermilgrb.html}}, we selected those GRBs detected in the energy range 30-100 MeV, namely 91 GRBs with available LLE data. 
We sorted these 91 GRBs according to the significance of their detection in the LLE energy range, which is measured by the parameter $\sigma$ using the Bayesian blocks detection algorithm, as reported in the LAT catalog. Since we are interested in investigating the high-energy part of GRB spectra, we required to have enough signal in the LLE data, namely we selected GRBs with $\sigma\geq20$.
Moreover, to ensure a good photon statistics also at lower energies, we required a fluence $F > 10^{-5}\,\rm erg\,cm^{-2}$ measured in the 10-1000\,keV energy range over the total duration of the burst, as reported in the GBM catalog. Applying these two criteria, we obtained 22 GRBs\footnote{We note that GRB 180113, despite matching our selection criteria (LLE detection with a $\sigma=40$ and a fluence $F = 1.63 \times 10^{-4}\,\rm erg\,cm^{-2}$), had no corresponding data in the LLE Catalog, thus it was not possible to include it in our final sample.}. 
All the selected events belong to the long GRB class.

\begin{figure}
\centering
\includegraphics[width=\columnwidth]{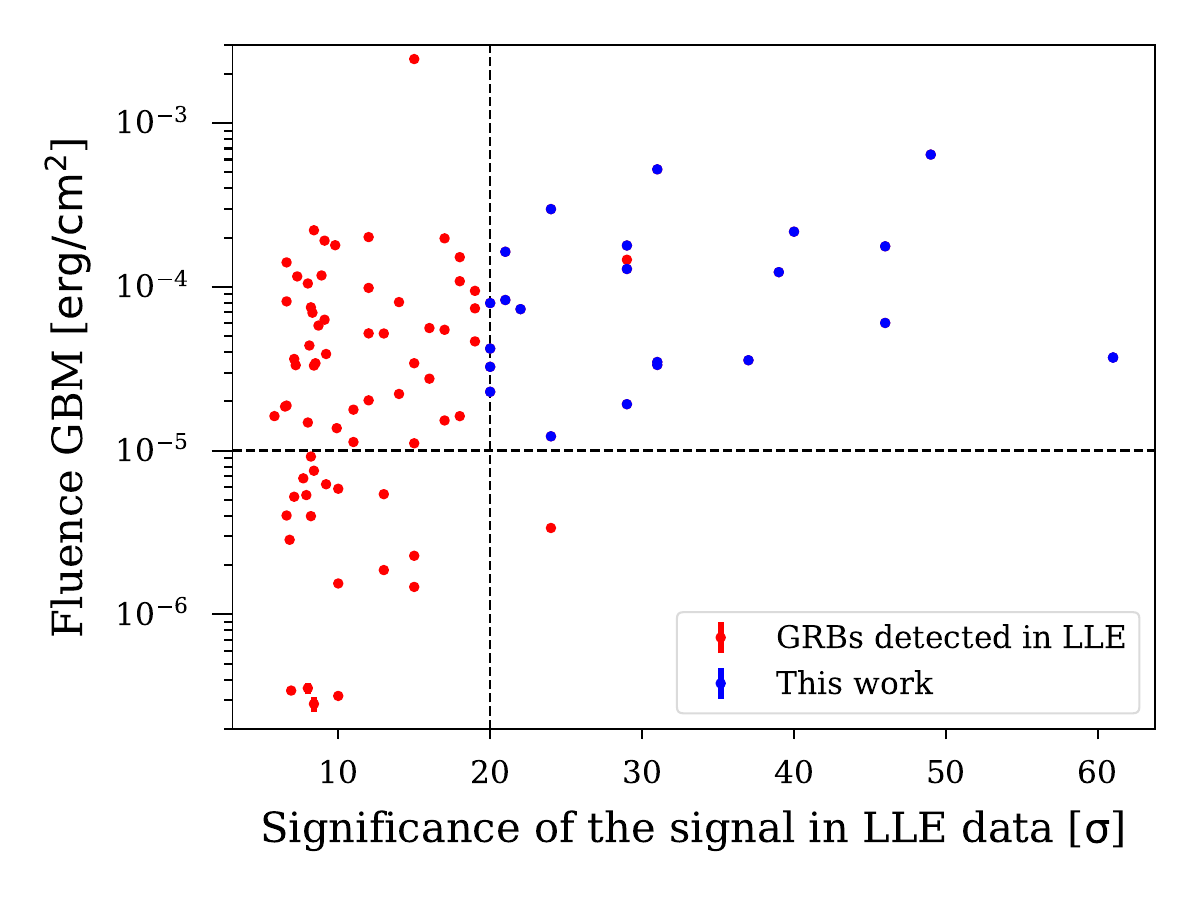} 
\caption{Sample of 22 GRBs analyzed (blue points) among the 91 events (red points) detected in the LLE energy range (30--100\, MeV), in 10 years of \fe\ activity, according to the LAT catalog. We selected the events applying a cut on the GBM fluence in the 10-1000 keV energy range and a cut on the significance of the signal in the LLE data, both represented by the two black dashed lines. The red point falling in the selected region represents GRB\,090926A, which has been excluded from the selection (see the text for an explanation). }
\label{fig:sampleGRB}
\end{figure}

Fig.~\ref{fig:sampleGRB} shows the fluence measured in the 10-1000\,keV energy range (i.e., detected by GBM) versus the significance of the signal in the range 30-100 MeV (i.e., LLE data), in units of $\sigma$, for the 91 GRBs of the LAT Catalog detected in the LLE energy range, represented by the red points. The blue points show the selected 22 GRBs. The selection criteria described above are represented by the two black dashed lines.
From our selection, we excluded GRB\,090926A (represented by the red point falling in the selected region in Fig.~\ref{fig:sampleGRB}). Conversely from the others, this GRB has an additional high-energy power-law component detected by the LAT during the prompt phase, which also shows a spectral cutoff at 1.4 GeV \citep{Ackermann2011,Yassine2017}. This prevents the identification of a possible high--energy cutoff on the main spectral component and the accurate determination of the slope of the high--energy power--law, which are both main goals of this work.

Table~\ref{tab:cutoff_sampleGRB} lists the name, 
the fluence in the 10-1000\,keV energy range and
the significance of the LLE signal or each selected event. We also list the redshift, if available 
(for 9 bursts).

\begin{table}[h]
\centering 
\caption{Sample of 22 long GRBs selected for this work. 
The events satisfy two selection criteria: a significance in the LLE [30--100 MeV] signal $\sigma \geq$ 20, and a fluence [10--1000\,keV] over the total duration of the burst $F > 10^{-5}\,\rm erg\,cm^{-2}$, as reported in the LAT and in the GBM catalog, respectively.
The first column lists the name of the GRB (with the last three digits referring to the naming convention of \fe\ GBM), the second column reports the fluence of the burst, the third column reports the significance of the signal in the LLE data, in terms of $\sigma$. The last column reports the redshift of the burst, if available. The events are listed in order of decreasing significance of the signal in the LLE data.}
\label{tab:cutoff_sampleGRB}
\footnotesize
\begin{adjustbox}{max width=\columnwidth}
\begin{tabular}{cccc}
\hline\hline
 \multicolumn{1}{c}{GRB Name} &
 \multicolumn{1}{c}{GBM Fluence} &
 \multicolumn{1}{c}{Significance of LLE signal} &
 \multicolumn{1}{c}{Redshift}\\
  & [$10^{-4}$\,erg\,cm$^{-2}$] & $\sigma$ & \\
\hline
110721[200] & 0.370 & 61 & --\\
160625[945] & 6.43 & 49 & 1.406\\
080916[009] & 0.603 & 46 & 4.35\\
170214[649] & 1.77 & 46 & 2.53\\
100724[029] & 2.17 & 40 & --\\
140206[275] & 1.23 & 39 & --\\
131108[862] & 0.357 & 37 & 2.4\\
141028[455] & 0.348 & 31 & 2.332\\
100116[897] & 0.334 & 31 & --\\
160821[857] & 5.22 & 31 & --\\
130504[978] & 1.29 & 29 & --\\
110328[520] & 0.192 & 29 & --\\
160509[374] & 1.79 & 29 & 1.17\\
180720[598] & 2.99 & 24 & 0.654\\
151006[413] & 0.122 & 24 & --\\
160905[471] & 0.732 & 22 & --\\
150902[733] & 0.832 & 21 & --\\
100826[957] & 1.64 & 21 & --\\
090328[401] & 0.420 & 20 & 0.736\\
110731[465] & 0.229 & 20 & 2.83\\
160910[722] & 0.797 & 20 & --\\
150202[999] & 0.325 & 20 & --\\
\hline
\end{tabular}
\end{adjustbox}
\end{table}

\section{Methods}\label{sect:methods}
For each GRB in our sample, we performed a joint spectral analysis of the GBM (10 keV -- 10 MeV), 
LLE (30 -- 100 MeV) and LAT (100 MeV -- 10 GeV) data (see shaded stripes in Fig.\ref{fig:gbm_lat_coverage}). In the following, we describe the data analysis procedure and the fitting models.

\subsection{Data analysis}

The GBM is an instrument composed of 12 sodium iodide (NaI, 8\,keV--1\,MeV) and two bismuth germanate (BGO, 200\,keV to 40\,MeV) scintillation detectors \citep{Meegan2009}, on board the \fe\ satellite. For each GRB, we analyzed the data from the two NaI detectors and the BGO detector with the highest count rates.
Spectral data files and the corresponding most updated response matrix files (rsp2) are obtained from the online archive \footnote{\url{https://heasarc.gsfc.nasa.gov/W3Browse/fermi/fermigbrst.html}}. We select the energy channels in the range 10--900\,keV for NaI detectors, and 0.3--10\,MeV for BGO detectors, and exclude the channels in the range 30--40\,keV due to the presence of the Iodine K-edge at 33.17\,keV\footnote{\url{https://fermi.gsfc.nasa.gov/ssc/data/analysis/GBM\_caveats.html}}. For our sample, which consists of long GRBs, we made use of CSPEC data, which have 1024\,ms time resolution.  Spectra were extracted with the public software {\sc gtburst}\footnote{\url{https://fermi.gsfc.nasa.gov/ssc/data/analysis/scitools/gtburst.html}}. 

The Large Area Telescope (LAT, \citealt{Atwood2009}) is a pair-conversion telescope covering the energy range from 30 MeV to more than 300 GeV. 
The extraction and analysis of LAT data were performed with {\sc gtburst}, following the procedure described in the online official Fermi guide \footnote{\url{https://fermi.gsfc.nasa.gov/ssc/data/analysis/scitools/gtburst.html}} and performing an unbinned likelihood analysis.
We selected P8R3\_TRANSIENT020 class events, filtering photons in the 100 MeV -- 10 GeV 
energy range from a region of interest (ROI) of 12$^\circ$ radius centered on the burst 
location. We applied a maximum zenith angle cut of 100$^\circ$ to reduce contamination 
of gamma-rays from the Earth limb.  
The LAT Low Energy (LLE) data were retrieved from the Fermi LLE catalog\footnote{\url{https://heasarc.gsfc.nasa.gov/W3Browse/fermi/fermille.html}} and reduced 
with a similar procedure as for the GBM data through {\sc gtburst}. 
The LLE spectra analyzed cover the energy range 30-100 MeV.

Spectral analysis is performed with the public software {\sc xspec}~(v.~12.10.1f). 
For the joint analysis of the GBM, LLE and LAT data, we used inter--calibration factors among the detectors, scaled to the brightest NaI and free to vary within 30\%.
We used the PG-Statistic, valid for Poisson data with a Gaussian background, in the fitting procedure.

For each GRB, we performed a time--integrated spectral analysis. 
Since this work aims at investigating the high-energy extension of the prompt spectrum, the time interval for the accumulation of the time-integrated spectrum for each GRB was selected based on the LLE light curve. The same time interval was adopted in extracting the time-integrated spectra from the GBM and LAT data.

\subsection{Fitting models}\label{subsec:fittingmodels}

\begin{figure}[h]
    \centering
    \includegraphics[width=\columnwidth]{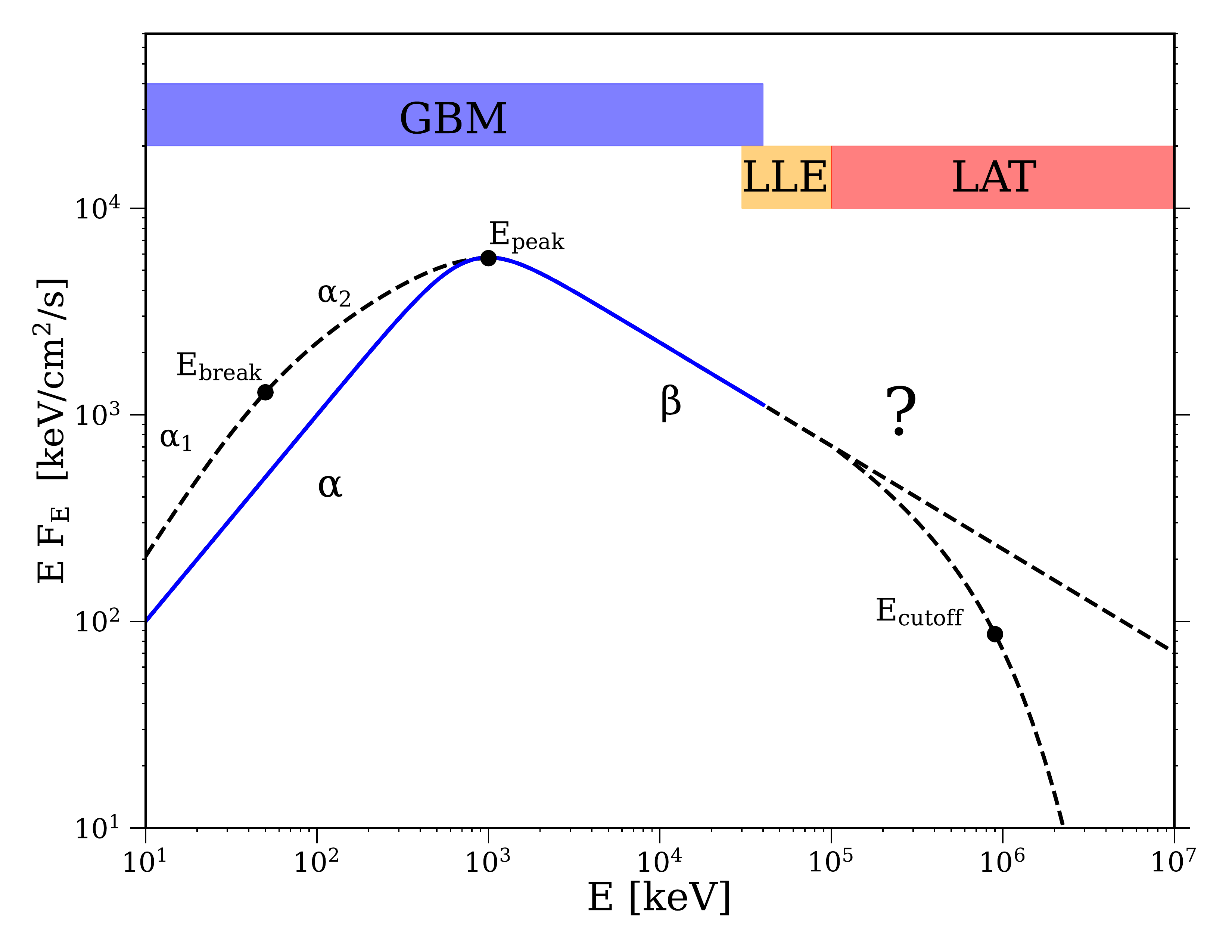}
    \caption{Representation of the four possible prompt emission spectral shapes and the corresponding spectral coverage of the GBM and the LAT instruments on board \fe\ tested in this work, spanning 6 decades in energy. The typical GRBs spectrum is shown with the blue solid line, characterized by the usual $\alpha$, $\beta$ and \ep\ parameters. Below the peak energy, the dashed black line shows the possible presence of a spectral break, modeled with the additional parameters $\alpha_{1}$, $\alpha_{2}$ and \eb\ (as shown in \citealt{Ravasio2018,Ravasio2019}). At higher energies, above \ep, the spectrum could show either the presence of an exponential cutoff, marked by $E_{\rm cutoff}$, or the extension of the power-law with slope $\beta$ (dashed black lines). 
    The y-axis scale is arbitrary.}
    \label{fig:gbm_lat_coverage}
\end{figure}

We analyzed each time--integrated spectrum with four different empirical functions: a smoothly broken power law (SBPL), a double smoothly broken power law (2SBPL, see \citealt{Ravasio2018} for the description of their functional form), and their modified versions, \textit{SBPLCUTOFF} and \textit{2SBPLCUTOFF}, respectively, which include the presence of a high--energy cutoff. Fig.~\ref{fig:gbm_lat_coverage} shows the four spectral shapes tested in this work, including the relative instrument we used to cover each energy range. The four fitting functions are described in the following. 

The SBPL is made of two power--laws, with spectral photon indices $\alpha$ and $\beta$, smoothly connected at a break energy (usually corresponding to the $\nu F_\nu$ peak of the spectrum, \ep). 
The 2SBPL is a function made of three power--laws (with photon indices $\alpha_1$, $\alpha_2$ and $\beta$) smoothly connected at two breaks (hereafter, \eb \,  and  \ep). 
The choice of fitting also with the 2SBPL is motivated by the fact that, 
besides searching for the presence of a cutoff at high energies, we also want to test the possibility of having a spectral break below the peak energy, as found in \citealt{Ravasio2018, Ravasio2019,Toffano2021, Gompertz2023}.

In order to identify the possible presence of a cutoff at high energies, we multiplied both SBPL and 2SBPL by an exponential cutoff function available among {\sc xspec} models, called \textit{highecut}. This piecewise function is defined as:

\[
  highecut(E) =
  \begin{cases}
                                   1 & \text{for $E \leq E_{\rm c}$} \\
                                   e^{(E_{\rm c} - E)/E_{\rm fold}} & \text{for $E \geq E_{\rm c}$}
  \end{cases}
\]

with $E_{\rm c}$ representing the energy at which the function starts to modify the basic model and $E_{\rm fold}$ regulating the sharpness of the exponential drop. The combination of the two quantities provides the energy at which the drop in the flux reaches a value of $1/e$, namely at $E_{\rm cutoff}= E_{\rm c}+E_{\rm fold}$. We consider $E_{\rm cutoff}$ as the relevant energy at which the spectrum shows a significant change with respect to the extrapolation of the power--law. 

The best-fit model is selected by minimizing the Akaike information criterion (AIC, \citealt{Akaike1974}). In particular, we select the more complex model only if the improvement of the fit corresponds to $\Delta$AIC $\geq 6$ \citep{BurnhamAIC} \footnote{The quantity $e^{-(AIC_{i} - AIC_{min})/2}=e^{-\Delta AIC/2}$ ($\sim$ 0.05 for $\Delta$AIC = 6) represents the relative likelihood of the i-th model with respect to the model with the lowest AIC value. It is proportional to the probability of the i-th model to minimize the information loss, namely the i-th model is $e^{-\Delta AIC/2}$ times as probable as the model with the lowest AIC to represent the data \citep{BurnhamAIC}.}.
The best--fit parameters and their uncertainties have been estimated through Markov Chain Monte Carlo (MCMC) approach.

\section{Results}\label{sect:results}

In Table~\ref{tab:cutoff_bestfitparams}, we present the results of the time-integrated analysis from 10 keV up to 10 GeV of the selected sample of LLE-detected GRBs.
For each GRB, the table lists the name, the best fit model (according to the $\Delta$AIC criterion reported above), the best fit parameters, the value of the PG-Statistic over the degrees of freedom, and, for those cases in which the spectrum is better fitted with the addition of a cutoff,  the corresponding $\Delta$AIC. The reported errors are at the 1-$\sigma$ level.
In the following sections, we focus on the evidence of an exponential cutoff at high energies and on the slope of the spectrum above the peak energy.

\begin{table*}
\small
\centering 
\caption{Best--fit parameters for the 22 long GRBs analyzed in this work. The first column displays the name of the burst, the second column reports the best--fit model, and the following columns report the corresponding best--fit parameters. If the best--fit model is the 2SBPL, the third and fifth columns show the indices $\alpha_1$ and $\alpha_2$ above and below the break energy \eb, which is reported in the fourth column. If the best--fit model is the SBPL, the photon index $\alpha$ is reported in the third column. 
The peak energy \ep\ and the high energy slope $\beta$ are reported for all best--fit models in the fourth and seventh column, respectively.
The cutoff energy, both as an estimate or as a lower limit, is reported in the eight column. GRBs showing a significant cutoff in the spectrum have been highlighted in boldface. The PG--Stat and DOF values are given for each GRB in column 9, while the corresponding improvement ($\Delta$AIC) of the fit with respect to the basic model (SBPL or 2SBPL) is reported in the last column only for GRBs showing a cutoff.}
\label{tab:cutoff_bestfitparams}
\begin{adjustbox}{max width=\textwidth}
\begin{tabular}{cccccccccc}
\hline\hline
  \multicolumn{1}{c}{Name} &
  \multicolumn{1}{c}{Best fit model} &
  \multicolumn{1}{c}{$\alpha_1 (\alpha)$} &
  \multicolumn{1}{c}{\eb} &
  \multicolumn{1}{c}{$\alpha_2$} &
  \multicolumn{1}{c}{\ep} &
  \multicolumn{1}{c}{$\beta$} &
  \multicolumn{1}{c}{$E_{\rm cutoff}$} &
  \multicolumn{1}{c}{PG--Stat/DOF} &
  \multicolumn{1}{c}{$\Delta$AIC} \\
  &  & & [keV] & & [keV] &  & [MeV] & &  \\
\hline
110721 & 2SBPL & ${-0.67}_{-0.07}^{+ 0.0}$ & ${59.7}_{-1.9}^{+ 16.3}$ & ${-1.51}_{-0.05}^{+ 0.01}$ & ${2394.9}_{-150.9}^{+ 215.1}$ & ${-2.75}_{-0.13}^{+ 0.01}$ & $\geq$689.6 & 361.1/319 & - \\
\textbf{160625} & \textbf{2SBPLCUTOFF} & $\mathbf{{-0.55}_{-0.01}^{+ 0.01}}$ & $\mathbf{{123.6}_{-4.3}^{+ 3.3}}$ & $\mathbf{{-1.71}_{-0.02}^{+ 0.02}}$ & $\mathbf{{738.1}_{-23.2}^{+ 20.7}}$ & $\mathbf{{-2.72}_{-0.03}^{+ 0.01}}$ & $\mathbf{{297.6}_{-30.6}^{+ 77.4}}$ & \textbf{655.7/352} & \textbf{12.5} \\
080916 & SBPL & ${-1.04}_{-0.02}^{+ 0.01}$ & - & - & ${545.9}_{-9.6}^{+ 60.6}$ & ${-2.23}_{-0.02}^{+ 0.04}$ & $\geq$ 3468.4 &  385.4/322 &  - \\
\textbf{170214} & \textbf{SBPLCUTOFF} & $\mathbf{{-0.87}_{-0.02}^{+ 0.01}}$ & - & - & $\mathbf{{363.6}_{-11.4}^{+ 12.0}}$ & $\mathbf{{-2.33}_{-0.04}^{+ 0.0}}$ & $\mathbf{{255.5}_{-49.9}^{+ 118.7}}$ &  \textbf{448.6/322} &  \textbf{22.6} \\
\textbf{100724} & \textbf{SBPLCUTOFF} & $\mathbf{{-0.82}_{-0.01}^{+ 0.01}}$ & - & - & $\mathbf{{603.7}_{-73.1}^{+ 88.4}}$ & $\mathbf{{-2.06}_{-0.03}^{+ 0.02}}$ & $\mathbf{{46.1}_{-5.1}^{+ 6.1}}$ &  \textbf{438.9/303} &  \textbf{167.7} \\
\textbf{140206} & \textbf{SBPLCUTOFF} & $\mathbf{{-0.98}_{-0.01}^{+ 0.01}}$ & - & - & $\mathbf{{395.2}_{-23.1}^{+ 29.1}}$ & $\mathbf{{-2.18}_{-0.03}^{+ 0.03}}$ & $\mathbf{{88.3}_{-12.5}^{+ 18.2}}$ &  \textbf{429.8/338} &  \textbf{69.1} \\
131108 & SBPL & ${-0.99}_{-0.01}^{+ 0.02}$ & - & - & ${366.2}_{-12.3}^{+ 12.3}$ & ${-2.23}_{-0.01}^{+ 0.03}$ & $\geq$ 825.1 &  471.9/431 &  - \\
141028 & 2SBPL & ${-0.84}_{-0.02}^{+ 0.04}$ & ${126.9}_{-9.6}^{+ 8.7}$ & ${-1.97}_{-0.01}^{+ 0.05}$ & ${1352.4}_{-92.4}^{+ 1428.3}$ & ${-2.96}_{-0.27}^{+ 0.07}$ & $\geq$488.5 & 346.3/319 & - \\
100116 & SBPL & ${-1.11}_{-0.01}^{+ 0.02}$ & - & - & ${702.6}_{-67.8}^{+ 64.6}$ & ${-2.69}_{-0.04}^{+ 0.06}$ & $\geq$ 502.2 &  458.3/414 &  - \\
\textbf{160821} & \textbf{2SBPLCUTOFF} & $\mathbf{{-0.77}_{-0.02}^{+0.04}}$ & $\mathbf{{136.2}_{-18.2}^{+39.2}}$ & $\mathbf{{-1.45}_{-0.04}^{+0.09}}$ & $\mathbf{{1615.8}_{-94.1}^{+56.3}}$ & $\mathbf{{-2.40}_{-0.04}^{+0.13}}$ & $\mathbf{{46.0}_{-4.4}^{+15.4}}$ & \textbf{487.5/319} & \textbf{193.2} \\
130504 & 2SBPL & ${-1.04}_{-0.02}^{+ 0.03}$ & ${208.0}_{-22.1}^{+ 13.0}$ & ${-1.96}_{-0.02}^{+ 0.06}$ & ${1377.7}_{-215.2}^{+ 110.4}$ & ${-3.34}_{-0.12}^{+ 0.21}$ & $\geq$86.5 & 587.5/382 & - \\
110328 & 2SBPL & ${-0.95}_{-0.1}^{+ 0.05}$ & ${131.6}_{-26.3}^{+ 96.8}$ & ${-1.74}_{-0.13}^{+ 0.06}$ & ${10559.9}_{-3710.5}^{+ 1162.3}$ & ${-4.6}_{-0.28}^{+ 0.95}$ & $\geq$128.6 & 360.3/300 & - \\
\textbf{160509} & \textbf{SBPLCUTOFF} & $\mathbf{{-0.89}_{-0.01}^{+ 0.01}}$ & - & - & $\mathbf{{385.6}_{-13.2}^{+ 22.2}}$ & $\mathbf{{-2.17}_{-0.02}^{+ 0.02}}$ & $\mathbf{{101.2}_{-17.3}^{+ 10.6}}$ &  \textbf{388.1/322} &  \textbf{171.1} \\
\textbf{180720} & \textbf{2SBPLCUTOFF} & $\mathbf{{-0.52}_{-0.09}^{+ 0.07}}$ & $\mathbf{{24.2}_{-1.6}^{+ 3.7}}$ & $\mathbf{{-1.36}_{-0.02}^{+ 0.01}}$ & $\mathbf{{788.1}_{-20.3}^{+ 42.1}}$ & $\mathbf{{-2.39}_{-0.05}^{+ 0.03}}$ & $\mathbf{{37.4}_{-4.3}^{+ 2.2}}$ & \textbf{517.8/318} & \textbf{42.2} \\
151006 & 2SBPL & ${-0.86}_{-0.11}^{+ 0.01}$ & ${50.3}_{-3.3}^{+ 34.7}$ & ${-1.67}_{-0.14}^{+ 0.03}$ & ${5319.3}_{-170.6}^{+ 5399.2}$ & ${-3.39}_{-0.64}^{+ 0.18}$ & $\geq$188.9 & 310.4/305 & - \\
\textbf{160905} & \textbf{SBPLCUTOFF} & $\mathbf{{-0.9}_{-0.02}^{+ 0.02}}$ & - & - & $\mathbf{{794.5}_{-59.0}^{+ 90.6}}$ & $\mathbf{{-2.4}_{-0.13}^{+ 0.12}}$ & $\mathbf{{14.3}_{-2.8}^{+ 4.7}}$ &  \textbf{398.1/303} &  \textbf{20.6} \\
150902 & 2SBPL & ${-0.58}_{-0.03}^{+ 0.03}$ & ${139.3}_{-11.1}^{+ 10.7}$ & ${-1.78}_{-0.08}^{+ 0.08}$ & ${368.7}_{-51.1}^{+ 39.5}$ & ${-2.82}_{-0.06}^{+ 0.04}$ & $\geq$495.3 & 379.5/320 & - \\
\textbf{090328} & \textbf{SBPLCUTOFF} & $\mathbf{{-1.08}_{-0.03}^{+ 0.02}}$ & - & - & $\mathbf{{651.3}_{-51.7}^{+ 63.0}}$ & $\mathbf{{-2.2}_{-0.05}^{+ 0.04}}$ & $\mathbf{{109.5}_{-6.3}^{+ 35.3}}$ &  \textbf{325.0/301} &  \textbf{25.8} \\
100826 & 2SBPL & ${-0.72}_{-0.04}^{+ 0.04}$ & ${112.7}_{-14.3}^{+ 15.6}$ & ${-1.74}_{-0.14}^{+ 0.08}$ & ${837.0}_{-137.9}^{+ 297.6}$ & ${-2.36}_{-0.21}^{+ 0.01}$ & $\geq$327.6 & 523.9/291 & - \\
110731 & SBPL & ${-1.04}_{-0.03}^{+ 0.02}$ & - & - & ${339.9}_{-12.5}^{+ 28.7}$ & ${-2.36}_{-0.01}^{+ 0.05}$ & $\geq$ 1941.9 &  380.7/320 &  - \\
160910 & 2SBPL & ${-0.49}_{-0.21}^{+ 0.03}$ & ${26.5}_{-0.1}^{+ 55.5}$ & ${-1.12}_{-0.19}^{+ 0.01}$ & ${332.1}_{-10.4}^{+ 22.0}$ & ${-2.36}_{-0.01}^{+ 0.04}$ & $\geq$202.4 & 422.8/302 & - \\
150202 & SBPL & ${-0.94}_{-0.04}^{+ 0.01}$ & - & - & ${216.8}_{-4.7}^{+ 12.1}$ & ${-2.46}_{-0.07}^{+ 0.02}$ & $\geq$ 56.7 &  388.6/302 &  - \\
\hline
\end{tabular}
\end{adjustbox}
\end{table*}

\subsection{High-energy spectral cutoff}

In nine out of 22 GRBs the addition of an exponential cutoff largely improves the fit, that is, their time-integrated spectrum shows a statistically significant ($\Delta \rm AIC \geq 20.6$, corresponding to $\gtrsim 4\sigma$) exponential drop at high energies with respect to the basic model. The cutoff energy $E_{\rm cutoff}$ is well constrained in all the 9 spectra and its distribution spans a wide range of values, from $\sim$ 14 up to 298 MeV.  In 4 GRB spectra, the cutoff energy is above 100 MeV, while for the remaining 5 it falls in the LLE energy range (i.e., $E_{\rm cutoff}$ is below 100 MeV).

\begin{figure}[h]
\centering
\includegraphics[width=\columnwidth]{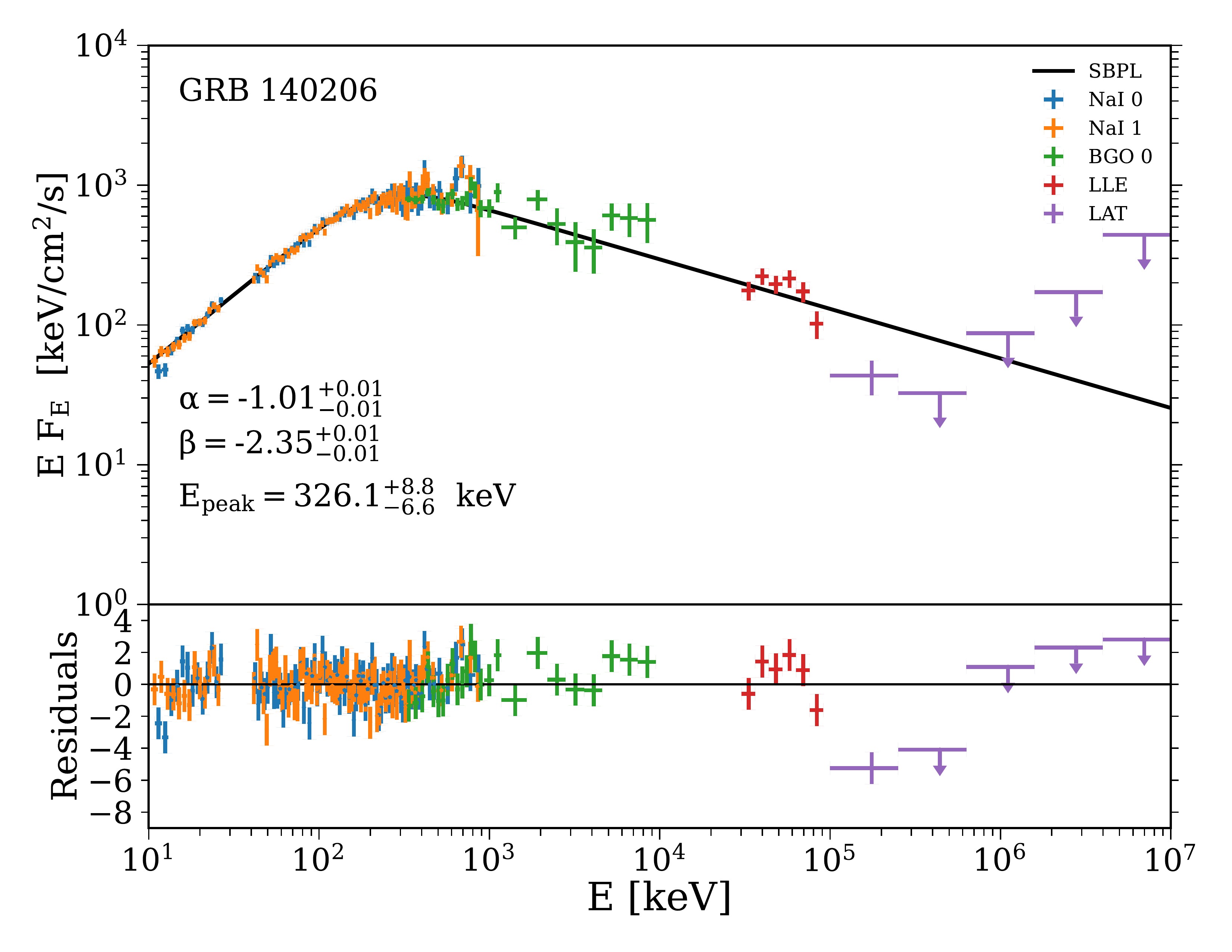} 
\includegraphics[width=\columnwidth]{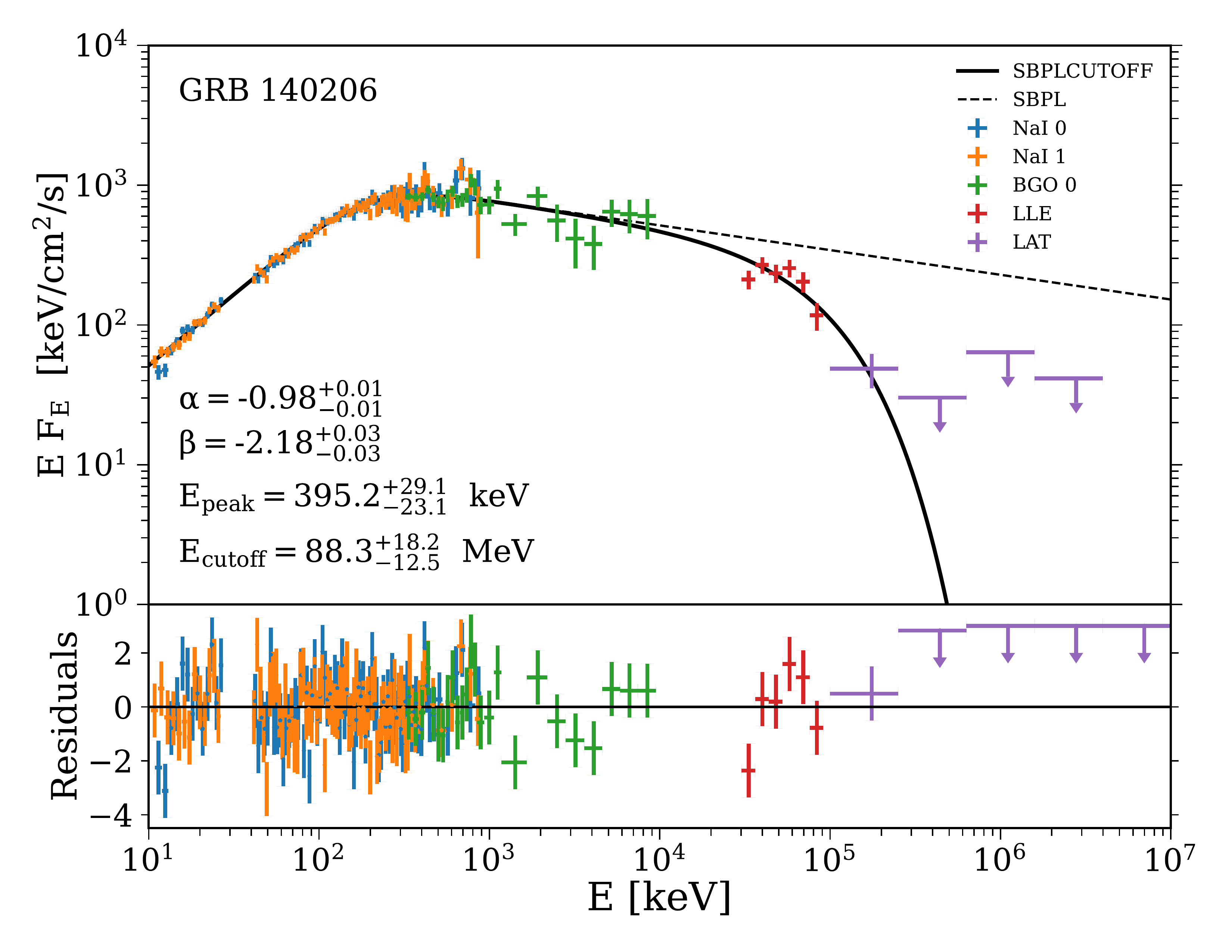} 
\vskip -0.2 cm
\caption{Example of a time-integrated spectrum showing a cutoff at high energy. Data correspond to the signal detected by different instruments (color-coded as shown in the legend) for GRB 140206. The top plot shows the fit with the SBPL function, represented by the solid black line. The bottom plot shows the same spectral data fitted with the addition of an exponential cutoff, namely with the SBPLCUTOFF function, represented with the solid black line. The dashed black line displays the SBPL function without the exponential cutoff term.
The parameters of the best--fitting models and their 1-$\sigma$ errors are reported in each plot (see also Tab.~\ref{tab:cutoff_bestfitparams}). The bottom panel of each plot shows the residuals of the spectral fit (computed as the distance of the data from the model in units of the errors). Data have been rebinned for graphical purposes.}
\label{fig:cutoff_140206}
\end{figure}

Fig.~\ref{fig:cutoff_140206} shows an example of a GRB spectrum showing a cutoff at high--energy. 
In particular, the data refer to the $\nu F_{\nu}$ representation of the time-integrated spectrum of GRB~140206.
The top plot shows the fit with the SBPL function (solid black line), which models the spectral shape with the following best--fit parameters: $\alpha = -1.01 \pm 0.01$ , $\beta = -2.35 \pm 0.01$ and $E_{\rm peak} = 326.1_{-6.6}^{+8.8}$ keV. However, both the LLE and LAT data are not properly modeled by the high--energy power--law with slope $\beta$, as the power--law overestimates the flux with respect to the data points from $\sim$ 80 MeV onward (see the residual panel reported below the plots).
The bottom plot shows the fit with the SBPLCUTOFF function. 
For comparison, the panel shows also the SBPL function without the exponential cutoff term (dashed black line). 
When fitted with the SBPLCUTOFF function, the spectral data are modeled by a larger peak energy $E_{\rm peak} = 395.2_{-23.1}^{+29.1}$ keV and a harder $\beta$ parameter $\beta = -2.18 \pm 0.03$ with respect to the fit with the SBPL. The exponential cutoff, parametrized by a cutoff energy at $E_{\rm cutoff} = 88.3^{+18.2}_{-12.5}$ MeV, allows to properly model the high-energy data and absorbs the previously visible trend in the residuals.
The addition of an exponential cutoff provides a highly significant improvement of the fit, statistically evaluated in this case by a $\Delta$AIC = 69.1. 

For 6 of the 9 GRBs showing a cutoff, the basic model that best describes the spectrum below the cutoff energy is the SBPL, while for the other 3 GRBs the best fit model is the 2SBPL. This means that for 3 GRBs, the spectrum from 10 keV to 10 GeV is characterized by 3 spectral breaks: the break energy \eb, the peak energy \ep\ and the cutoff energy $E_{\rm cutoff}$. This result shows how crucial is a wide and continuous energy coverage of the GRBs prompt emission spectrum, when possible, to systematically characterize the whole spectral shape and unveil possible deviations from the typically adopted fitting functions (Band and SBPL), carrying potentially valuable physical information about the emitting region.\\

For the remaining 13 GRBs (out of 22), the presence of a spectral exponential cutoff at high energies is not supported by the data, thus the high--energy spectrum is best fitted with a single power--law up to 10 GeV. 
Since the high--energy data are often characterized by large error bars, they can possibly hide the presence of an exponential cutoff, which may be allowed by the fit statistic above a given energy. 
Therefore, for the 13 GRBs not showing a cutoff, we aimed at setting a lower--limit on the position of  $E_{\rm cutoff}$. 
The procedure adopted is the following. We fit the data with the basic model (SBPL or 2SBPL) and then fixed the parameters to the found best--fitting values. Then we added a cutoff at progressively smaller energies\footnote{To perform these tests we used the command \textit{steppar} implemented in XSPEC, which performs a fit while stepping the value of a parameter through a given range.}, until the fit statistic allows to consider the addition of the cutoff as statistically equivalent to the model without the cutoff. Using the same criterion adopted for choosing the best--fitting model, the presence of a cutoff is no longer supported by the data when it leads to a $\Delta$AIC$\geq$6 with respect to the best-fit statistic. The value of the cutoff energy $E_{\rm cutoff}$ corresponding to the first fit with $\Delta$AIC$\geq$6 is considered as the lower--limit of the cutoff energy for the analyzed spectrum.
This procedure led to obtaining the 13 lower--limits on $E_{\rm cutoff}$, reported in Table~\ref{tab:cutoff_bestfitparams}, for the corresponding 13 GRBs not showing a cutoff.

\subsection{Beta distribution}
For all the models tested, the shape of the spectrum above the peak energy is modeled by a power--law with slope $\beta$. This allows us to study the high--energy spectral slope for each GRB of our sample, regardless of the presence of a cutoff. 

\begin{figure}[h]
    \centering
    \includegraphics[width=\columnwidth]{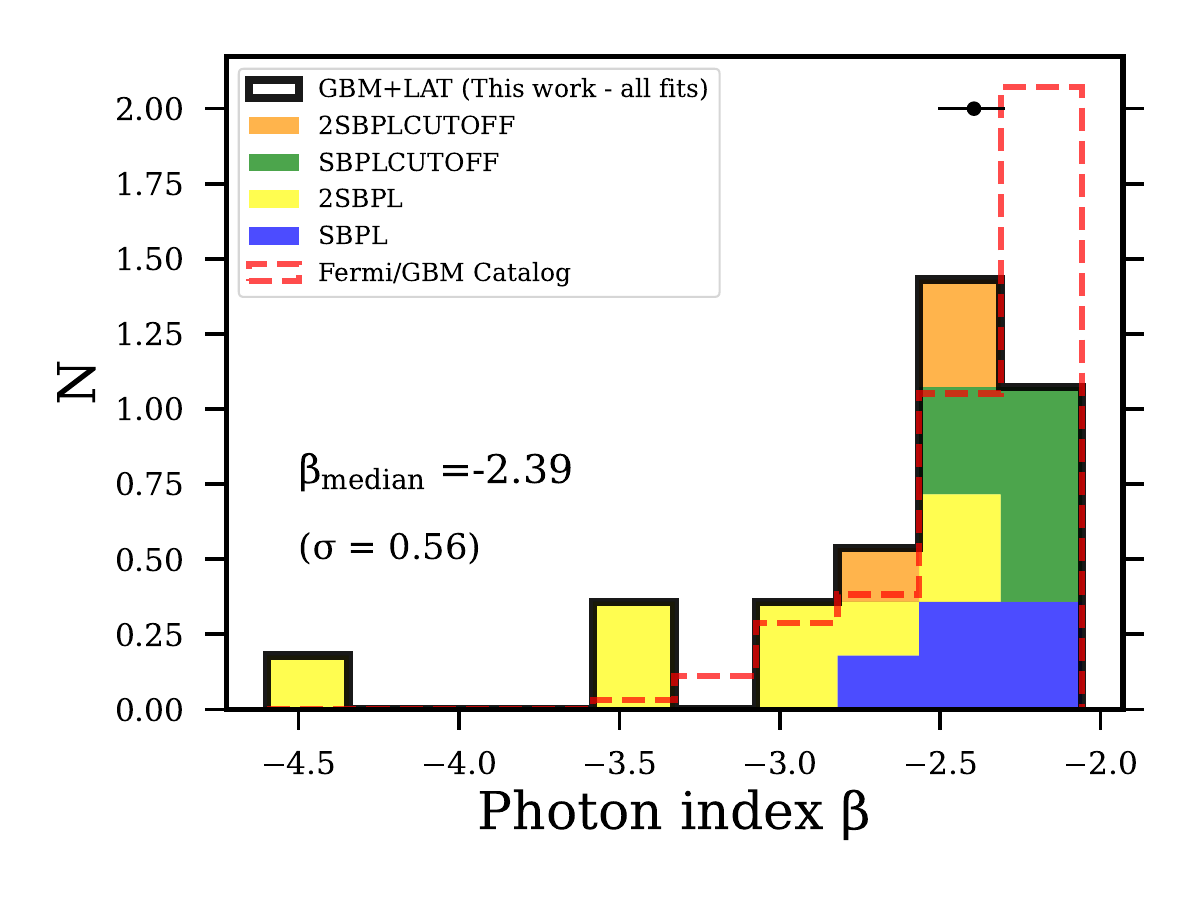}
\vskip -0.3 cm
\caption{Distributions of the high-energy slope $\beta$ (photon index) for the 22 GRBs' spectra analyzed. The different best--fitting models are represented by different colors (stacked on top of each other), while the black empty histogram represents the total distribution of $\beta$ values derived from the spectral fits performed in this work. On the top of the distributions it is over--plotted the median value and the mean error on the $\beta$ parameter for the whole sample. For comparison, the distribution of the $\beta$ values derived from the \fe/GBM catalog is represented by the red dashed histogram. The distributions have been normalized to the respective sample size. 
    }
    \label{fig:beta_distrib}
\end{figure}

By combining the results from all the best--fitting models shown in Table~\ref{tab:cutoff_bestfitparams}, we built the distribution of the $\beta$ parameter, 
which we reported with the black empty
histogram in Fig.~\ref{fig:beta_distrib}. It is important to note that the distribution does not show values greater than $-2$. 
This is simply due to the hard limit we set on the $\beta$ parameter,
in order to have a peak in the $\nu F_{\nu}$ spectrum.
All the spectra analyzed in our sample showed a well-defined peak in the broad energy band studied, such that, if the constraint on the $\beta$ parameter would be relaxed, we would still have $\beta$ values smaller than $-2$.

The distribution of $\beta$ shown in  Fig.~\ref{fig:beta_distrib} spans a wide range, from -4.60 to -2.06, and it is has a median value of 
$\langle \beta \rangle$ = -2.39 (standard deviation of 0.56).
The average error on the $\beta$ values together with the median value of the distribution are represented by the black horizontal bar and the point over-plotted above the histograms.

In this plot, the yellow and blue histograms represent the $\beta$ parameter when the best--fitting models are 2SBPL and SBPL, respectively. The orange and green histograms display the $\beta$ values of the spectra best--fitted by the 2SBPLCUTOFF or the SBPLCUTOFF model, respectively.
Comparing the $\beta$ values of the spectra best--fitted by the SBPL with the ones best fitted by the 2SBPL, we note that the latter ones tend to be steeper than the first ones 
(see the blue and the yellow histograms in the plot). This confirms the results found in previous studies \citep{Ravasio2018,Ravasio2019}, and it is likely due to the addition of the low energy break, that allows the 2SBPL function to better model the peak energy and thus the remaining high--energy part of the spectrum.

We now compare the distributions of $\beta$ for the 13 GRB spectra not showing a cutoff (yellow and blue histograms in Fig.~\ref{fig:beta_distrib}) 
with the 9 GRBs showing a cutoff (orange and green distributions). 
The median value of $\beta$ for the 13 GRB spectra not showing a cutoff is 
$\langle \beta_{\rm no\, cutoff} \rangle = -2.69$ ($\sigma$= 0.64), 
while for the 9 GRBs showing a cutoff is 
$\langle \beta_{\rm cutoff} \rangle = -2.33$ ($\sigma$= 0.19). 
The small difference between the two distributions could be explained as an effect due to the presence of the exponential cutoff. In fact, under the assumption that the intrinsic high--energy slope is the same, when an exponential cutoff is present in the spectrum, the power--law segment between the peak energy and the cutoff energy is truncated earlier, allowing to constrain $\beta$ on a smaller energy range. This effect, combined with the broad curvature of the spectrum, likely provides shallower (i.e., harder) values of the high--energy slope of GRBs showing a cutoff. \\

We now compare these results on the high-energy power--law slope $\beta$ with the whole Fermi GBM catalog. 
Among the four functions used to fit the spectra in the catalog, only the Band and the 
SBPL ones provide a $\beta$ parameter for the modeling of the high-energy part of the spectrum. 
Due to the necessity of fitting a large variety of spectral shapes, in the Fermi
catalog no constraints on the spectral indices of both the Band and SBPL functions are introduced. Therefore, differently from our results, the \fe/GBM distribution $\beta$ extends beyond values of $\beta= -2$ ($\sim$18\% of the values are greater than $-2$, at more than 1$\sigma$). 
When excluding values greater than $-2$ from the \fe/GBM catalog distribution, we find that the $\beta$ distribution obtained in this work is characterized by steeper (i.e., softer)
values than the one reported in the catalog. 
In Fig.~\ref{fig:beta_distrib}, we show with the dashed red histograms the distribution of the $\beta$ slopes of the \fe/GBM  catalog bursts whose time--integrated spectrum has been best--fitted by Band or SBPL function, and with $\beta < -2$: the median value of this distribution $\langle \beta_{\rm Fermi} \rangle$ is $-2.29$ ($\sigma=$ 0.33).  
A K-S test returns a $p$-value=0.022\footnote{We recall here that for the statistical tests, we have set the significance level at 0.05, i.e., we accept the null hypothesis if $p>0.05$.}, suggesting ($\sim2.3\sigma$) that the two distributions might not be drawn from the same parent distribution. 

The diversity in the $\beta$ distributions could be ascribed to the addition of the LLE and LAT data to the GBM ones in our analysis. While in the \fe/GBM catalog are reported the parameters of the fit obtained considering spectral data up to 40 MeV, in this analysis we extended the characterization of the spectra up to 10 GeV. 
This may suggest that high-energy slope of the GRBs prompt emission spectrum is indeed softer than what is commonly deduced from fits made with lower energy data alone (e.g., from BATSE or GBM).
A similar conclusion has been drawn also by \citealt{Ackermann2012}, where it was shown that in 24 bursts co-detected by GBM and LAT the $\beta$ values are systematically softer than the values found from fitting the GBM data alone. Also, when performing the joint fit of GBM and LAT data, they found no cases of spectra with $\beta > -2$, in agreement with the constraint we set on this parameter.
Perhaps less dominant, but still of impact on our $\beta$ distribution, could be also the effect of the use of more complex fitting functions in our analysis, as compared to the simpler Band and SBPL function used in the catalog. 

\section{Interpretation}\label{sect:interp}

In this section, we provide the interpretation of the results obtained from the analysis of the high-energy extension of 22 GRBs bright prompt emission spectra, described in the previous section. 
This section is divided into two parts. The first part regards the evidence of the 
exponential cutoff, which will be interpreted as the sign of pair-production opacity, allowing us to derive the bulk Lorentz factor of the jet. The alternative interpretation, involving the maximum photon energy for synchrotron emission, will be discussed in Sec.~\ref{sect:discussion}.
The second part of this section concerns the results on the slope 
$\beta$ of the high-energy power--law, which allows to put valuable
constrain on the slope of the underlying distribution of the accelerated particles.

\subsection{Estimate of $\Gamma$ from pair-production opacity}

By interpreting  the observed exponential cutoff as due to the absorption caused by the photon--photon process, producing electron--positron pairs, we can infer an estimate of the bulk Lorentz factor $\Gamma$ of the jet during the prompt emission phase
\citep[e.g.,][]{Piran1999, lithwick01}.

We now derive a computation of the optical depth for pair--production in the emitting region, which depends on the spectral parameters derived from the analysis in Sec.~\ref{sect:results}, 
and use it to set a lower limit or an estimate of $\Gamma$, depending on whether an exponential cutoff is absent or present in the spectrum, respectively. 
Primed quantities are calculated in the comoving frame.

For a photon with energy $h \nu$, the energy $h \nu_{\rm T}$ of the target photons for the $\gamma \gamma \xrightarrow{} e^{+}e^{-}$ process is such that $\epsilon^{\prime}_T = 1/\epsilon^{\prime}$, where $\epsilon^{\prime} = h\nu^{\prime}/(m_e c^2)$ and $\epsilon^{\prime}_{\rm T} = h\nu^{\prime}_{\rm T}/(m_e c^2)$.
Following Equation B3 reported in \citealt{Svensson1987}, we can express the optical depth for pair-production for a photon of energy $\epsilon^{\prime}$ as:
\begin{eqnarray}
\tau_{\gamma \gamma} (\epsilon^{\prime}) = \eta(\beta_e) \sigma_T \frac{1}{\epsilon^{\prime}} n(\frac{1}{\epsilon^{\prime}}) \Delta R^{\prime} =  \eta(\beta_e) \sigma_T \frac{U_{rad}^{\prime}(1/\epsilon^{\prime})}{m_e c^2} \Delta R^{\prime}
\label{taugammagamma}
\end{eqnarray}

where $\sigma_T$ the Thomson cross-section, $\Delta R^{\prime}$ the width of the shell 
and $U_{\rm rad}^{\prime}$ the radiation energy density. 
We are assuming that the spectrum of the target photons is a power law of energy index $\beta_e$,
(not to be confused with the $photon$ spectral index $\beta$).
Therefore the comoving luminosity, at these energies, is 
$L^\prime(\epsilon_T^{\prime})\propto (\epsilon_T^\prime)^{-\beta_e}$. 
As a consequence, also the monochromatic radiation energy density 
${U_{rad}}^{\prime}(\epsilon_T^{\prime}) \propto (\epsilon_T^{\prime})^{-\beta_e}$, 
and the photon number density 
$ n^{\prime}( \epsilon_T^{\prime} ) = {U_{rad}}^{\prime}(\epsilon_T^{\prime})/\epsilon_T^{\prime}
\propto ( \epsilon_T^{\prime} )^{-(1+\beta_e)}$.

Eq. \ref{taugammagamma} mainly depends on the number density at threshold and weakly on the
spectral shape of the number photon density. 
This is because of the steep decrease of the cross-section 
$\sigma_{\gamma\gamma} (\epsilon_T^{\prime})$ with energy.
Furthermore, for typical spectral indices $\beta_e$, also 
 $n^{\prime}(\epsilon_T^{\prime})$ decreases with energy.
 \citealt{Svensson1987} numerically calculated the correct $\tau_{\gamma\gamma}$
encapsulating the spectral index dependence into the function $\eta(\beta_e)$ which
can be approximated with (see his Eq. B5) the following:
\begin{equation}
\eta(\beta_e) = {7\over 6}\, {1\over (2+\beta_e) (1+\beta_e)^{5/3} }
.\end{equation}
For the range of $\beta_e$ of interest in our calculations (1--1.5) 
we have $\eta(\beta_e)$=0.122--0.072.

Assuming the usual expression that links the width of the shell $\Delta R^{\prime}$ to the distance from the central engine $R$, namely $\Delta R^{\prime} = R/\Gamma$,
and rewriting the optical depth as a function of the frequency $\nu$,
we obtain:

\begin{eqnarray}
\tau_{\gamma \gamma} (\nu) =  \eta(\beta_e) \sigma_T \frac{L^{\prime}(\nu_T)}{4\pi R^2 c h} \frac{R}{\Gamma}.
\label{eq:taugg_R}
\end{eqnarray}

Here, we use the assumption that $R=2ct_{\rm var}\Gamma^2$, where $t_{\rm var}$ is the variability timescale, and apply the relativistic transformation $L^{\prime}(\nu_T)=L(\nu_T)/\Gamma$,   
obtaining:
\begin{eqnarray}
\tau_{\gamma \gamma} (\nu) = \eta(\beta_e) \sigma_T \frac{d_{L}^{2} F(\nu_T)}{2 c^2 h t_{\rm var} \Gamma^4} 
\end{eqnarray}

Since the spectrum of the target photons is assumed to be a power--law with spectral energy index $\beta_e$, we can rewrite  $F (\nu_T) = F (\nu_{peak}) \, \left(\frac{\nu_T}{\nu_{peak}}\right)^{-\beta_e}$, finding:

\begin{eqnarray}
\tiny
\tau_{\gamma \gamma} (\nu_{\rm cutoff}) = \eta(\beta_e) \sigma_T d_{L}^{2}  F(\nu_{\rm peak}) \left[\frac{(m_e c^2)^{2}}{h \nu_{\rm peak}}\right]^{-\beta_e} \frac{h \nu_{\rm cutoff}^{\beta_e}}{2 c^2 h t_{\rm var}} \Gamma^{-4-2\beta_e}
\end{eqnarray}

This equation would allow to compute the optical depth for every GRB whose parameters  $F(\nu_{\rm peak})$, $\beta_e$, $\nu_{\rm peak}$, $\nu_{\rm cutoff}$, $t_{\rm var}$ and $\Gamma$ are known. However, instead of using $F(\nu_{\rm peak})$, namely the monochromatic flux at the peak energy $h \nu_{\rm peak}$ of the spectrum, we can further rewrite this equation using more common parameters for GRBs spectra, such as the isotropic luminosity $L_{\rm iso}$ and the spectral photon indices of the Band function $\alpha$ and $\beta$, converted in energy indices $\alpha_e$ and $\beta_e$. Hence, if the spectrum is described by two power--laws with energy indices $\alpha_e$ and $\beta_e$ connected at the peak energy $h \nu_{\rm peak}$, we can express the optical depth 
as:

\begin{equation}
\begin{aligned}
\tiny
\tau_{\gamma \gamma} (E_{\rm cutoff}) = \eta(\beta_e) \sigma_T \frac{L_{\rm iso}}{4\pi} \frac{(1-\alpha_e)(\beta_e-1)}{(\beta_e-\alpha_e)} \times \\
\times \left[\frac{(m_e c^2)^{2}}{E_{\rm peak}}\right]^{-\beta_e} \frac{E_{\rm cutoff}^{\beta_e}}{2 c^2 E_{\rm peak} t_{\rm var}} \Gamma^{-4-2\beta_e}
\label{eq:cutoff_taugg}
\end{aligned}
\end{equation}

The requirement that $\tau_{\gamma \gamma}(E_{\rm cutoff}) = 1$ in Eq.~\ref{eq:cutoff_taugg} 
leads to the estimate of $\Gamma$:

\begin{equation}
\tiny
\Gamma = \left[\eta(\beta_e) \sigma_T \frac{L_{\rm iso}}{4\pi} \frac{(1-\alpha_e)(\beta_e-1)}{(\beta_e-\alpha_e)} \left[\frac{(m_e c^2)^{2}}{E_{\rm peak}}\right]^{-\beta_e} \frac{E_{\rm cutoff}^{\beta_e}}{2 c^2 E_{\rm peak} t_{\rm var}}\right]^{{\frac{1}{4+2\beta_e}}}
\label{eq:cutoff_Gamma} 
\end{equation}

For the computation of the bulk Lorentz factor, we need 
the redshift $z$ of the source, which is required to compute the isotropic luminosity, the rest frame energies and the variability timescale in the equation.
For 9 GRBs in our sample, the redshift has been spectroscopically measured (see Table \ref{tab:cutoff_sampleGRB}), and we used it in the computation of $\Gamma$. 
For those GRBs without a redshift measurement, we used $z=2$ as representative value of the redshift distribution of long GRBs\footnote{We repeated the analysis assuming four different redshift values, i.e., $z$ $\in$ [0.5, 1 , 3, 4], finding that the corresponding values of $\Gamma$ are on average a factor [0.50, 0.70, 1.25, 1.48] times the ones derived assuming $z=2$. The choice of not testing redshift values higher than $z=4$ is motivated by the extreme values of the corresponding $E_{\rm iso} (\gtrsim 4 \times 10^{55}$ erg) otherwise implied by higher redshift for the considerably bright GRBs selected in this work.}.
Each GRBs light curve shows different temporal behaviors and varies on different timescales. 
For the sake of simplicity, we assume the same value of the variability timescale for all the 
GRBs, namely $t_{\rm var}$ = 0.1 s,  which corresponds to the mean value of the variability 
timescales for long bursts observed by Fermi \citep{Golkhou2015}.

Setting $\tau_{\gamma \gamma}(E_{\rm cutoff}) = 1$ in Eq.~\ref{eq:cutoff_taugg} allows us 
to derive an estimate of $\Gamma$ for those 9 GRBs showing a cutoff in their spectrum, 
as reported in Sec.~\ref{sect:results}. 
For the remaining 13 GRBs not showing a cutoff, we used the lower--limit on $E_{\rm cutoff}$ 
derived from the spectral fits in order to set a lower--limit on their $\Gamma$. 
The uncertainties on the measured quantities $E_{\rm cutoff}$, $E_{\rm peak}$, 
$L_{\rm iso}$, $\alpha_{\rm e}$ and $\beta_{\rm e}$ give a relative uncertainty 
on the order of $\sim 4$\% on the estimates of $\Gamma$.
These values of $\Gamma$ must be  compared with $\Gamma_{\rm max} = (1+z) \, E_{\rm cutoff}/m_e c^2$. This limit corresponds to the maximum bulk Lorentz factor attainable for a given observed 
cutoff energy, since the cutoff energy in the comoving frame would be at the self-annihilation 
threshold ($E_{\rm cutoff}^{\prime} = m_e c^2$) \footnote{This is due the fact that, when we 
observe a cutoff energy in the spectrum, it could correspond at least
to $m_e c^2$ in the 
comoving frame, which directly gives a limit to the maximum value of bulk Lorentz factor 
$\Gamma$.}. 
The bulk Lorentz factor is then the minimum of the obtained values.
We note that, for all the 9 GRBs showing a cutoff, the estimate of $\Gamma$ is below the computed maximum value $\Gamma_{\rm max}$: we therefore consider the value derived from Eq.~\ref{eq:cutoff_taugg} as the estimate of their bulk Lorentz factor.

\begin{figure}[h]
    \centering
    \includegraphics[width=\columnwidth]{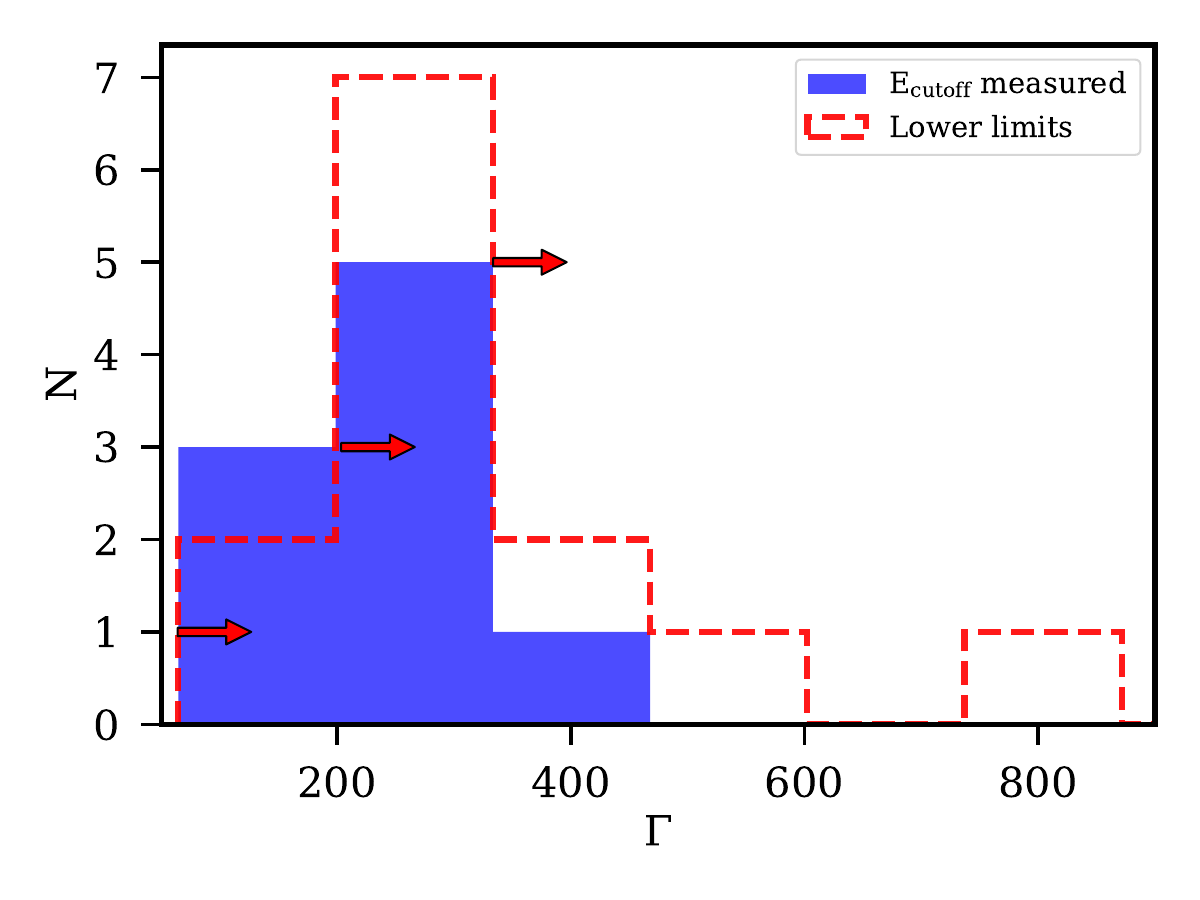}
    \vskip -0.3 cm
    \caption{Distribution of bulk Lorentz factor $\Gamma$ obtained from the spectral analysis of the 22 bright GRBs that were co-detected by GBM and LAT and considered for this work. The values of $\Gamma$ are derived by interpreting the high--energy cutoff (either a measurement of or a lower--limit of) as a sign of pair-production opacity, following Eq.~\ref{eq:cutoff_Gamma}.  
    The blue-filled histogram represents the measurement of $\Gamma$ for the 9 GRBs of the sample showing evidence of high--energy cutoff in their prompt emission spectra.
    The dashed red histogram (with rightward arrows) represents the lower--limit on $\Gamma$ from the 13 GRBs not showing a cutoff in their spectrum (see the text for details).
    }
    \label{fig:Gamma0}
\end{figure}

Table~\ref{tab:cutoff_R} shows the values and the lower--limits of $\Gamma$ derived from Eq.~\ref{eq:cutoff_Gamma} for each GRB in our sample, either with or without the detection of an exponential cutoff in the spectrum.
The distribution of the estimates of $\Gamma$ for the 9 GRBs showing a cutoff is shown with the blue filled histogram in Fig.~\ref{fig:Gamma0}, together with the distribution of the lower--limits on $\Gamma$ for the 13 GRBs not showing a cutoff (red dashed histogram). The distribution of measured $\Gamma$ values is centered around $\langle \Gamma \rangle = 229 \, (\sigma=68)$, while the distribution of the lower--limits on $\Gamma$ has a mean value of $\langle \Gamma \rangle = 317 \, (\sigma=157)$. The main reason for the difference in the two distributions can be due to the fact that, for the 13 GRBs without the detection of a cutoff, it has been used the lower--limit on the cutoff energy to derive $\Gamma$ and their values are on the order of several hundreds of MeV (typically a factor $\sim$ 5--6 times larger than the estimates of $E_{\rm cutoff}$ for the 9 GRBs with a detected exponential cutoff). The lower--limits on $E_{\rm cutoff}$ imply that there could be a cutoff at higher energies. Speculating on the presence of such high-energy cutoffs, which would not be observable with the current instruments, this would in turn imply that the distribution of $\Gamma$ 
would be broader, extending toward larger values. 
This is indeed compatible with what inferred from the afterglow observations, as it will be discussed in Section~\ref{sec:comparison_w_afterglow}.\\

\subsection{Estimate of p} 

From the sample of 22 GRBs co-detected by GBM and LAT, we derived the distribution of the slope $\beta$ of the high-energy power--law above \ep, shown in Fig.~\ref{fig:beta_distrib}. 
The spectral photon index $\beta$ is the key parameter to derive the spectral index $p$ of the population of shock--accelerated particles.

Electrons are assumed to be injected continuously in the emitting region, according to a unique power law (of index $p$) between $\gamma_m$ and $\gamma_{max}$. 
In the framework of the synchrotron interpretation of GRB spectra, these energies correspond to the synchrotron frequencies $\nu_m$ and $\nu_{max}$. 
The slope of the emitted radiation depends on the cooling regime. 
In fast cooling regime, the spectrum is described by $F_{\nu} \propto -p/2$ for $ \nu_{\rm m} < \nu < \nu_{\rm max}$, while in slow cooling regime $F_{\nu} \propto -(p-1)/2$ for $\nu_{\rm m} < \nu < \nu_{\rm c}$.

\begin{figure}[h]
    \centering
    \includegraphics[width=\columnwidth]{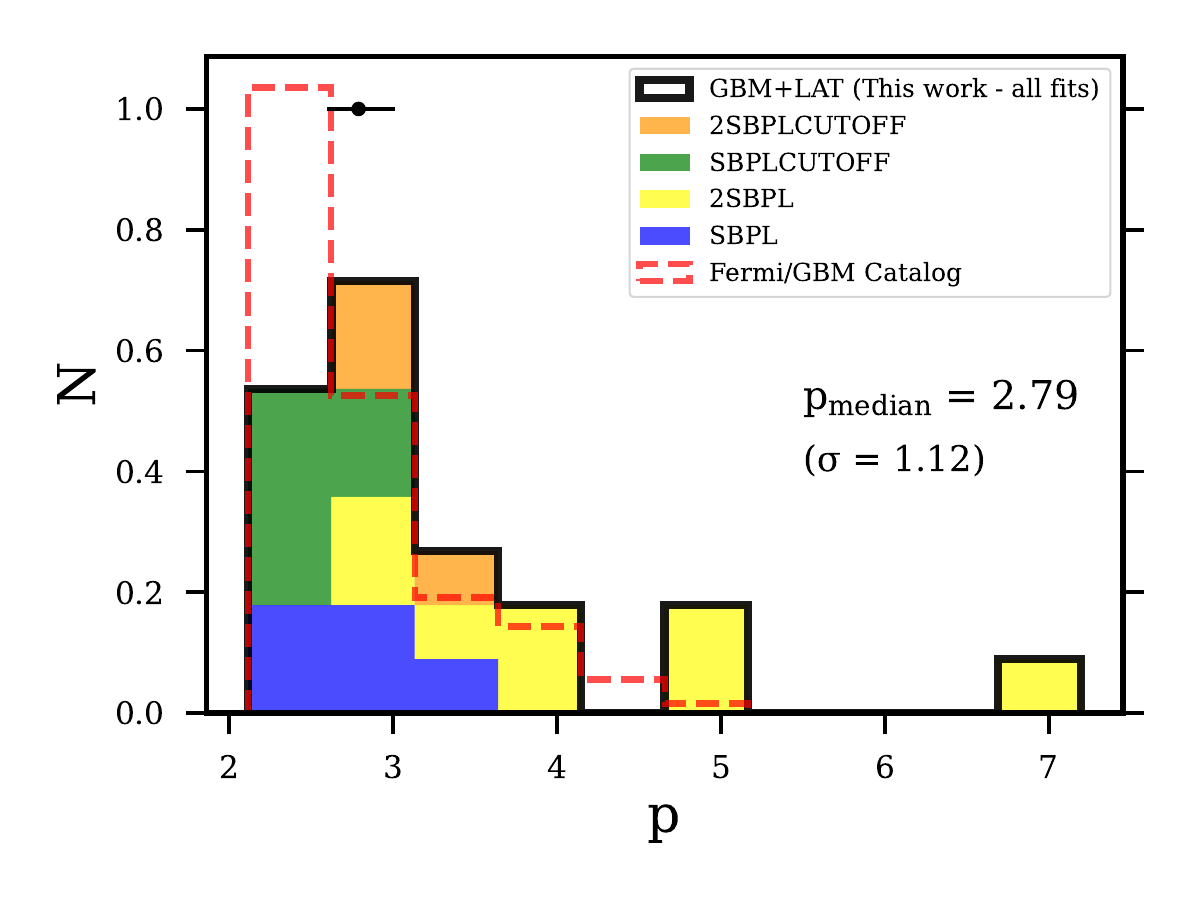}
    \vskip -0.4 cm
    \caption{Distribution of the slope $p$ of the nonthermal population of accelerated 
    particles for the sample of 22 GRBs analyzed in this work. This distribution has been derived from the $\beta$ distribution shown in Fig.~\ref{fig:beta_distrib}, 
    by using the relation $p = -2 -2\beta$ 
    or, equivalently, $p=2\beta_e$
    (i.e., assuming synchrotron emission in the fast cooling regime). 
    The median value of $p$ along with its error bar on $p$ values 
    are represented with the black dot and solid line, respectively.
    For comparison, the distribution of $p$ derived for the \fe/GBM catalog bursts under the same assumption is also plotted (red dashed histogram).}
    \label{fig:p_distrib}
\end{figure}

In 11 out of 22 GRBs analyzed, the spectrum shows the presence of two characteristic energies, \eb\, and \ep\ (with \eb < \ep), which we associate to the synchrotron cooling frequency $\nu_{\rm c}$ and the injection frequency $\nu_{\rm m}$, respectively, thus showing a fast cooling regime. 
Given the previous results on the brightest bursts detected by \fe/GBM \citep{Ravasio2019}, showing spectra with $\nu_{\rm  c} < \nu_{\rm m}$, and on the simulations performed in \citet{Toffano2021}, demonstrating that observational biases (e.g., detector threshold and fluence) can prevent the cooling break identification even when this is indeed present in the spectrum, we will assume that all the 22 GRBs spectra analyzed are in fast cooling regime. 
Under this assumption, we associate each spectral slope $\beta_e$ (where $\beta_e =\beta-1$) to the slope of the accelerated particle distribution using the relation $p$ = 2$\beta_e$ 
\footnote{If, on the contrary, the spectrum of those GRBs not showing a break is in slow cooling regime, the corresponding 
$p$ values would be greater than the ones in fast cooling regime, i.e., $p_{\rm slow\, cooling} = p_{\rm fast \,cooling} +1$, pushing even more to the right the $p$ distribution in Fig.~\ref{fig:p_distrib}.}.

Fig.~\ref{fig:p_distrib} shows the distribution of $p$ resulting from the $\beta$ distribution derived from the 22 GRBs analyzed (shown in Fig.~\ref{fig:beta_distrib}). 
Each color represents the distribution of the slope obtained from the corresponding spectral model fitted to the data, while the black empty histogram represents the total distribution.
Combining all the fit results, the distribution of $p$ has a median value of $p$ = 2.79, with a standard deviation of 1.12.
As done for the $\beta$ distribution, we compare our results with the \fe/GBM catalog.
In the same Fig.~\ref{fig:p_distrib}, we plotted with the red dashed histogram the distribution of $p$ derived from the $\beta$ values reported in the \fe/GBM catalog bursts (for those best--fitted by Band or SBPL function and with $\beta < -2$), and assuming synchrotron in fast cooling regime.
As noted for the $\beta$ distribution, also the values of $p$ obtained from our analysis tend to be steeper than the one derived from the \fe/GBM catalog.
This can be due to the fact that we studied the extension of the spectrum at higher energies with respect to the GBM, adding crucial data from 30 MeV up to 10 GeV that help to better constrain the high--energy slope of the spectrum.

There are two main considerations that we can draw from these results. The first one regards the width of the $p$ distribution.
Fig.~\ref{fig:p_distrib} shows that the distribution of $p$ is quite broad, with a $\sigma=1.12$. The majority of $p$ values are clustered between 2 and 3.5, but there is also a non-negligible tail extending toward steep values. This result may suggest that the acceleration mechanism does not produce relativistic particles distributed as a power--law with a unique slope, but rather with a quite wide variety of slopes. 
A similar conclusion has been drawn for bursts detected by BATSE \citep{Shen2006}, where the observed distribution of $p$, centered around $\langle p \rangle = 2.50$ with $\sigma=0.49$, was found to be inconsistent with a universal value.

The second consideration regards the steepness of the $p$ slopes. In fact, the 
distribution of $p$ shows values generously extending at $p\sim$ 5 (up to 7 in one case).
Such steep values of $p$ have been found also in other works in the literature. 
As an example, the synchrotron modeling of the prompt emission 
spectrum of GRB~180720B shown in \citet{Ronchi2020} returned a slope $p=4.8$ of the injected population of electrons (see their Fig. 5, SED I [0-35 s]).
This in turn produced a soft photon spectrum  ($\beta  \sim -3.4$) at high energies, as confirmed also by the fit of the high-energy spectrum with a single power law. 
Such steep values of $p$ set strong constraints on the accelerations mechanism, either internal shocks (\citealt{Bednarz1998,Achterberg2001,Lemoine2003}, see \citealt{Sironi2015b} for a recent review) 
or magnetic reconnection
\citep{Spruit2001, Drenkhahn2002}, \citep{Sironi2014, Sironi2015a, Petropoulou2019}.
Indeed, the acceleration mechanism responsible for the injection of nonthermal electrons must be efficient in accelerating electrons to high energies, but it also should give rise to a steep electron energy distribution.


\section{Discussion}\label{sect:discussion}

\subsection{Comparison with previous detections of exponential cutoff}

We here compare our results with earlier published works that found 
the presence of an exponential cutoff in
the prompt emission spectrum. For this comparison, we exclude those works reporting a cutoff 
in the extra power--law components in the spectrum (e.g., see \citealt{Ackermann2010,Ackermann2011,Ackermann2013,Abdo090902B,Abdo2009_080916C}), as we focus on the main prompt emission component, typically represented by the peaked Band-like spectral shape.

\citealt{Ackermann2012} analyzed 288 GRBs detected by GBM that do not show evidence of emission above 100 MeV. 
They compared the computed flux upper limits in the LAT 
energy band with the extrapolations of the spectral fits of GBM data, finding that 6 GRBs of the sample required the presence of a spectral cutoff between GBM and LAT data. 
However, lacking a direct evidence of a spectral cutoff, they set $E=100$ MeV as the cutoff energy for all 6 bursts.

\citealt{Tang2015} systematically searched for a high-energy spectral cutoff 
combining GBM and LAT data of 28 GRBs prompt emission spectra. 
They fit the time-integrated prompt spectrum with the Band 
function with and without a high--energy cutoff. 
They found five GRBs showing a cutoff at energies between $\sim$13 and $\sim$62 MeV. 
Three of the five GRBs reported to have a cutoff are also 
present in our sample. 
Only in one of them, namely GRB~140206, we also found evidence of a spectral cutoff  with  $E_{\rm cutoff}=88.3^{+18.2}_{-12.5}$ MeV
which is higher than the value  $50.1 \pm$6.8 MeV reported in \citep{Tang2015}. 
For the remaining 2 GRBs, we do not find any evidence of a cutoff and our best--fitting model is the 2SBPL. 
These discrepancies could be ascribed to the use of different fitting models (Band function in their analysis vs the set of SBPL/2SBPL in our work), although a more quantitative comparison should be performed to support this explanation.

GRB~160625B, included in our sample, has been reported by \citealt{Wang2017} 
to have an exponential cutoff of tens of MeV in six out of eight time-resolved intervals. 
This is consistent with the time-integrated result reported also in \citealt{Ravasio2018} 
and with the results obtained in this work. 

\citealt{Vianello2018} studied the prompt emission spectra of GRB~100724B and GRB~160509A, 
both included in our sample. 
Their time-resolved analysis revealed the significant presence of high-energy 
cutoff at $E_{\rm cutoff} \sim$ 20--60 MeV for GRB~100724B and $\sim$80--150 MeV for GRB~160509A, with $E_{\rm cutoff}$ slightly increasing with time in both cases. 
We found consistent results from the time-integrated analysis on similar time 
intervals, reporting a $E_{\rm cutoff}={45.4}_{-5.6}^{+ 12.2}$ MeV for GRB~100724B and $E_{\rm cutoff}={100.7}_{-16.8}^{+ 9.9}$ MeV for GRB~160509A.

In \citealt{Ryde2022}, the authors used synchrotron spectral fits of the prompt emission in GRB 160821, finding good agreement with the data.  
They found that the spectral shape is characterized by three spectral breaks, 
namely at $h\nu_{\rm c} = 167.4_{-12.4}^{+14.2}$ keV, 
$h \nu_{\rm m} = 2.3_{-1.0}^{+1.1} 10^3$ keV and 
$h \nu_{\rm cutoff} = 5.2_{-1.7}^{+1.9} 10^4$ keV. 
These results are consistent with our findings, which show that GRB 160821 
is best--fitted by the 2SBPLCUTOFF function, and in particular the values 
of the spectral breaks found in their analysis are consistent with the ones 
presented in this work (see Table~\ref{tab:cutoff_bestfitparams}), 
despite the different models tested (synchrotron model vs. empirical 
function, respectively). 
However, they do not interpret the cutoff energy as due to the opacity to pair-production, 
but rather as the high-energy cutoff included in their synchrotron model. 

Combining all the previous detections reported in the literature, the systematic analysis of high-energy spectra reported in this work revealed 
4 new cases of GRB prompt spectra showing the presence of an exponential cutoff: GRB~170214, GRB~180720, GRB~160905, and GRB~090328.

\subsection{Comparison with estimates of $\Gamma$ from pair-production opacity in the literature}

We now focus on the comparison of the estimates of $\Gamma$ with those derived 
in literature for other GRBs using the same pair-production opacity argument. 

In the work of \citealt{Tang2015}, the computation of pair-production opacity $\tau_{\gamma \gamma}$ is similar to the one in this work and also to the one presented in \citet{lithwick01}, namely assuming a simple one--zone model where the photon field in the emitting region is uniform, isotropic, and time-independent. However, since the cutoff energies in the main spectral components they reported were below $\sim$ 60 MeV, in the majority of the cases their estimates of $\Gamma$ exceeded the maximum attainable values $\Gamma_{\rm max}$. This led the authors to use $\Gamma_{\rm max}$ as the actual bulk Lorentz factor for the majority of the bursts analyzed, finding values of $\Gamma$ between $\sim$ 50 and $\sim$150. These values are systematically lower than the values inferred in this work. 

For the remarkable cases of GRB~100724B and GRB~160509A, showing cutoff energies from $\sim$ few tens of MeV up to 150 MeV, \citet{Vianello2018} used two different time-dependent models to derive the values of the bulk Lorentz factors (i.e., an internal--shocks motivated model by \citealt{Granot2008} and a photospheric model by \citealt{Gill2014}). The authors reported bulk Lorentz factors in the range $\Gamma \sim$ 100 -- 400, which are consistent with our results.

It is important to note that simplified one-zone models, as the one used in this work and in \citet{lithwick01}, \citet{Tang2015} and \citet{Ackermann2012}, may provide systematic differences in the inferred Lorentz factors as compared to more detailed time-dependent multi-zone models, where multiple emitting regions and the time-, space- and direction-dependent photon field are taken into account. \citet{Granot2008}, and more recently \citet{Hascoet2012} and \citet{Gill2018}, have shown that such time-dependent models can yield inferred $\Gamma$ estimates that are reduced by a factor of $\sim$2 compared to estimates made using single-zone models. In the context of these time-dependent models, the estimates of $\Gamma$ presented in Fig.~\ref{fig:Gamma0} should all be rescaled downward by a factor of $\sim$2. Moreover, \citet{Granot2008}, investigating the opacity effects for impulsive emission between a given radius $R_0$ and $R_0 + \Delta R$, found that the time-integrated spectrum should display a power-law tail (for impulsive emission with $\Delta R/R_0 \lesssim 1$)  
instead of an exponential cutoff (for $\Delta R/R_0 \gtrsim 1$ - see also the more recent work of \citealt{Dai2023}) above a given cutoff energy. 
However, the signal-to-noise of spectral data in the MeV-GeV range typically offered by the current instruments makes it challenging to distinguish a break into a steep power-law from an exponential cutoff at high energies, although a direct comparison of these two models (broken power-law and exponential cutoff) on actual spectral data should be performed
to investigate this. In the work of \citet{Vianello2018}, where a spectral cutoff has been found in GRB 100724B and GRB 160509A using an empirical function (similar to what is done in this work) and also the model described in \citealt{Granot2008}, the authors tested whether a power-law after the cutoff energy provide a better fit than the exponential, and concluded that, despite the increased complexity, this is never the case, for both GRBs, suggesting that the shape of the spectrum after the cutoff appears to be curved.
For a discussion of single and multi-zone models, see also \citet{Zou2011}.\\

We note here that an alternative interpretation of the observed exponential cutoff would be the maximum frequency for synchrotron emission, which reflects the maximum energy $\gamma_{\rm max}$ of the shock-accelerated particles distribution \citep{Guilbert1983}. 
The maximum energy $\gamma_{\rm max}$ is directly connected to the shock micro-physics conditions and it is typically estimated by equating the timescale for synchrotron cooling and the acceleration timescale. It is expected at much higher energies with respect to the cutoff energies found in this work, both for the electrons 
($E^{\prime \rm max, syn}_{\rm e} \sim$ 50 MeV in the rest frame) and for the protons ($E^{\prime \rm max, syn}_{\rm p} \sim$ 93 GeV, see \citealt{Ghisellini2020}). We stress that this value is a limit and therefore, in principle, the exponential cut-off energies observed in this analysis could be explained as the maximum synchrotron frequency, provided that 
the distribution of accelerated particles spans a quite limited energy range from $\gamma_{\rm min}$ (given by \ep) to $\gamma_{\rm max}$ (corresponding to the cutoff energy $E_{\rm cutoff}$) and that the radiative losses are due to synchrotron emission.
Under this interpretation, our results on the cutoff energies would provide only lower limits on the bulk Lorentz factor of the jet.

\subsection{Comparison with $\Gamma$ derived from the afterglow onset time}\label{sec:comparison_w_afterglow}

The majority of the estimates of the bulk Lorentz factor $\Gamma$ for GRBs reported in the literature have been inferred from the measurement of the time of the peak in the afterglow light curve \citep{Molinari2007,Liang2010,Lu2012,Ghirlanda2012,Ghirlanda2018}.
In the standard fireball scenario, such a peak is commonly interpreted as due to the transition from the coasting (when the jet travels with a constant bulk Lorentz factor) to the deceleration phase (when the jet starts to decelerate due to the interaction with the interstellar medium), pinpointing the onset of the afterglow.

In this section we focus on the comparison of the estimates of $\Gamma$ obtained using two different physical processes, pair-production opacity (this work) and deceleration of the jet. For the latter, we compare with the work of \citealt{Ghirlanda2018}, which reported the values of $\Gamma$ derived from the systematic study of the early afterglow light curve of the "gold sample" of \citep{Ghirlanda2018}.

\begin{figure}[h]
    \centering
    \includegraphics[width=\columnwidth]{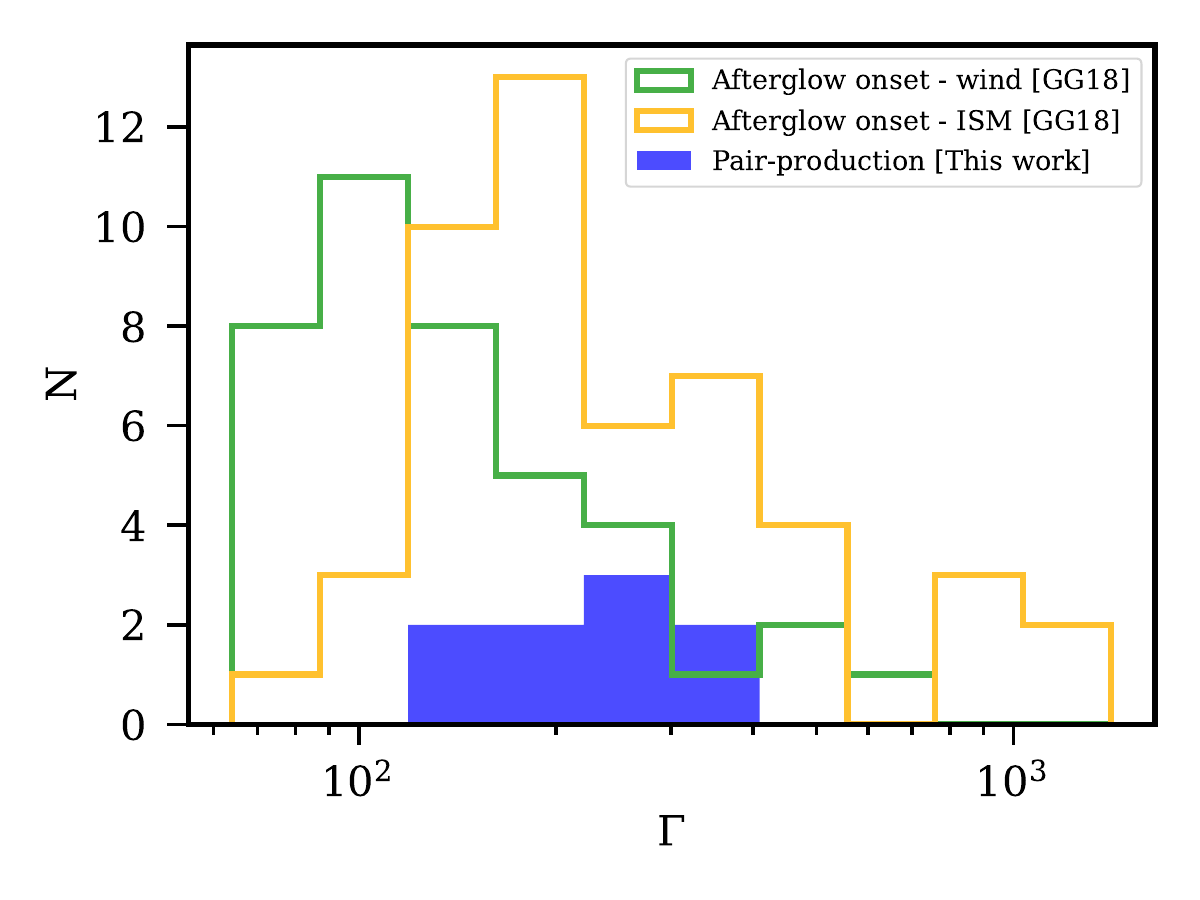}
\vskip -0.3 cm
    \caption{Comparison of the distributions of bulk Lorentz factors of GRBs obtained from independent methods. The blue-filled histogram represents the distribution of $\Gamma$ values derived in this work from the pair-production opacity argument applied to the prompt spectra of 9 GRBs showing a cutoff. The empty histograms show the distributions of $\Gamma$ values derived from the measurement of the afterglow onset time for a sample of 50 GRBs, as reported in \citealt{Ghirlanda2018}. The orange histogram represents the values of the bulk Lorentz factor derived assuming a constant medium density, while the green histogram refers to the wind--like medium case. These distributions are built taking into account measurements only (not lower limits), for both methods.}
    \label{fig:cutoff_gamma0distrib}
\end{figure}

Fig.~\ref{fig:cutoff_gamma0distrib} shows the distributions of the values of the bulk 
Lorentz factors inferred from the two methods (taking into account only the measurements and not the lower--limits). 
The filled blue histogram represents the estimates of $\Gamma$ derived from the $\gamma \gamma$ opacity argument applied in this work, while the empty histograms represent the values derived from the measurement of the afterglow onset time as reported in \citealt{Ghirlanda2018}, with the orange (green) color representing the homogeneous (wind-like) medium. 
For the homogeneous medium case, the authors assumed a typical medium density of $n_{\rm 0} = 1$ cm$^{-3}$. 
For the wind-like case, they used a normalization of $n_{\rm 0} = 10^{35} \dot{M}_{-5} \, v_{\rm w, 3}^{-1}$ cm$^{-3}$, following the relation 
$n(r) = n_{\rm 0} r^{-2}= \dot{M} /(4 \pi m_{\rm p} v_{\rm w} r^2)$, where typical values of the mass-loss rate $\dot{M} = 10^{-5} M_{\odot}$ yr$^{-1}$ 
and of the wind velocity $v_{\rm w} = 10^{3}$ km s$^{-1}$ have been assumed
\citep{Chevalier2000}.
We stress, however, that we are comparing the distributions assuming that all GRBs in \citealt{Ghirlanda2018} have an ISM medium (orange distribution) or a wind (green distribution) medium, while of course for each burst only one medium type, and by extension only one value of $\Gamma$, is relevant, and a detailed analysis of the individual broad-band afterglow emission should be performed to distinguish the adequate medium for each burst.

Overall, the values of $\Gamma$ derived in this work are consistent with those derived 
from the afterglow onset reported in \citet{Ghirlanda2018}. 
Our results are well included in the 68\% of the distribution of 
$\Gamma$ derived from the afterglow onset time assuming a homogeneous medium. 
A K--S test performed between the constant ISM case and pair--production $\Gamma$ 
distributions returns a $p$-value = 0.64, indicating that they are likely drawn 
from the same underlying distribution of bulk Lorentz factors.
Our results are instead located in the higher tail of the distribution related to the wind--like medium, and a K--S test returns a $p$-value = 0.003.
Since in the wind case the assumed density of the circumburst medium is higher, 
the $\Gamma$ values related to the wind--like case are typically smaller than 
the ones related to the homogeneous case. Keeping fixed all the other parameters 
in Eq.~\ref{eq:cutoff_Gamma}, smaller values of $\Gamma$ would imply smaller 
values of the cutoff energy $E_{\rm cutoff}$, which in turn implies that the high energy prompt emission
may not be detected in the LAT energy band. 
In turn, this may indicate a selection effect in favor of faster 
GRBs when analyzing GRBs detected by \fe/LAT. 
On the other hand, there may be a bias also on the right side of 
the distribution, namely toward higher values of the bulk Lorentz factor. 
The distribution of $\Gamma$ may extend toward higher values, 
as indeed already suggested by the distribution of lower-limits on $\Gamma$ (red dashed histogram in Fig.~\ref{fig:Gamma0}), which extends up to $\Gamma\sim$750. However, keeping fixed all the other parameters of 
Eq.~\ref{eq:cutoff_Gamma}, such higher values of $\Gamma$ imply higher values of $E_{\rm cutoff}$, which cannot be observed with the current instruments.

Since the $\Gamma$ values derived from the afterglow onset time depends on 
the medium density, this comparison may be affected by a different choice 
for this parameter (and for the wind velocity and mass loss rate). 
For example, in order to reproduce the $\Gamma$ distribution found in our analysis 
(blue filled histogram), the distribution derived from the afterglow onset for the 
wind case (green empty histogram) should be moved toward higher values of 
$\Gamma$ and this implies to assume smaller densities of the external medium.
This is more relevant for the wind case ($\Gamma \propto n^{-1/4}$)
than in the case of homogeneous ISM ($\Gamma \propto n^{-1/8}$).

\subsection{Estimate of the radius $R$}

GRB~080916 and GRB~110731 belong to both our sample and the sample of 
GRBs with afterglow onset time measured in \citet{Ghirlanda2018}.
Therefore, for both GRBs one could perform a direct comparison of the bulk Lorentz factors 
derived from the two independent methods.

From our analysis, their prompt emission spectra do not show the presence of a cutoff, as they are both best--fitted by the SBPL. 
Therefore, we can only set a lower--limit on their $\Gamma$ values from the 
pair--production argument. In 
Table~\ref{tab:cutoff_R}, for GRB 080916 and 110731 we report the bulk Lorentz factors derived from the prompt cutoff measure and from the afterglow light curve peak (the latter both considering a homogeneous and a wind--like medium scenario). 
The bulk Lorentz factor $\Gamma$ derived from the afterglow onset assuming a homogeneous 
medium is in agreement with the lower--limit implied by the pair-production argument 
derived in this work. 
Instead, the value derived for the wind--like medium case is smaller than our lower--limit.

On one hand, from the afterglow perspective the values for the density of the circumburst 
medium may play a significant role in the derivation of $\Gamma$ for the wind--like 
medium, as noted before, while on the other hand, from the pair--production argument perspective, 
we are considering the internal-shock framework and assuming a variability timescale.
In deriving the bulk Lorentz factor reported in Sec.~\ref{sect:interp}, we set 
$t_{\rm var}$ = 0.1 s for all GRBs, based on the average value of the observed 
distribution of variability timescale for a sample of \fe\, detected 
GRBs \citep{Golkhou2015}. 
For every GRBs of our sample, the implied value of the radius $R$ (following the usual $R= 2 c t_{\rm var} \Gamma^2/(1+z)$) is reported in the first row of the fourth column of the Table~\ref{tab:cutoff_R}.

However, this value may likely not correspond to the intrinsic value of the variability timescale for each burst in our sample and this may affect the estimated $\Gamma$ and, consequently, also the value of the distance $R$ of the GRB emitting region from the central engine. The distance $R$ is one of the least known and often assumed (via the variability timescale) parameters of the GRB standard model.
Indeed, as pointed out in \citealt{Gupta2008} and used in for example \citealt{Zhang2009} and \citealt{Chand2020}, a thorough treatment free from the assumptions on the variability timescale and on the internal shocks would be to build a two-dimensional constraint in the $\Gamma - R$ plane.
To break the degeneracy, we can then take advantage of having the bulk Lorentz factor constrained from an independent method (e.g., the afterglow light curve), when available, and constrain the radius $R$. 

Indeed, one can invert Eq.~\ref{eq:taugg_R} (see also Eq.(13) in \citealt{Gupta2008}) and use the $\Gamma$ value inferred from the afterglow onset to derive the value (or a lower--limit) of the radius $R$, applying the compactness argument, without making assumptions on the variability timescale. 

Solving Eq.~\ref{eq:cutoff_taugg} for the radius $R$, we obtain:

\begin{equation}
\tiny
    R = \eta(\beta_e) \sigma_T \frac{L_{\rm iso}}{4\pi} \frac{(1-\alpha_e)(\beta_e-1)}{(\beta_e-\alpha_e)} \left[\frac{(m_e c^2)^{2}}{E_{\rm peak}}\right]^{-\beta_e} \frac{E_{\rm cutoff}^{\beta_e}}{c E_{\rm peak}} \Gamma^{-2-2\beta_e} .
    \label{eq:cutoff_R}
\end{equation}

This method has the advantage of being independent of the assumptions on (or the measurements of) the variability timescale $t_{\rm var}$. 
It is always possible, however, to compute the variability timescale a posteriori for a given $\Gamma$ and $R$: indeed, in the last column of Table~\ref{tab:cutoff_R} we report the variability timescales implied by the corresponding values of $\Gamma$ and $R$, when this method has been applied.

If the GRB spectrum shows an exponential cutoff, then the cutoff energy $E_{\rm cutoff}$  allows to derive the estimate of $R$ (by setting $\tau_{\gamma \gamma} (E_{\rm cutoff}) = 1$, as before). 
If instead, the spectrum is a single power--law above the peak, then the lower--limit on the cutoff energy will allow us to derive only a lower--limit on $R$.
We stress, however, that this method holds under the assumption that the bulk Lorentz factor $\Gamma$ of the jet derived from the afterglow onset corresponds to the average value of the distribution of bulk Lorentz factors of the jet during the prompt emission. If instead, the jet would have a  $\Gamma$ at the deceleration higher than the average value of $\Gamma$ of the prompt emission (e.g., due to magnetic acceleration, as suggested in Poynting-flux-dominated models), then this assumption would not hold.

We show in Table~\ref{tab:cutoff_R} the relative estimate of $R$, for GRB~080916 and GRB~110731, following Eq.~\ref{eq:cutoff_R} and using the values of $\Gamma$ obtained for the homogeneous and for the wind--like medium by \citealt{Ghirlanda2018}. 
The two estimates of the radius derived in the two cases are quite different: $R$ in the wind--like case is greater by two orders of magnitude 
than in the homogeneous case. 
This is due to the corresponding $\Gamma$ values, which differ by a factor of $\sim$2--3.
As shown in Eq.~\ref{eq:cutoff_R},  $R \propto \Gamma^{-2-2\beta_e}$: for a typical energy index 
$\beta_e \sim 1.5$, we have $R \propto \Gamma^{-5}$.
In both cases, however, the values of the distance $R$ are found to be $R \gtrsim 10^{13-15}$ cm.
This might suggest that, although still consistent with it, the distance of the emitting region from the central engine could be slightly larger than the radius typically assumed in the GRB standard model for the internal shocks to occur. We note that large radii ($R \sim 10^{15}$ cm) would be favored in Poynting-flux dominated models, such as the Internal-Collision-induced Magnetic Reconnection and Turbulence (ICMART) model \citep{Zhang2011b}, in order to reach the condition of triggering a reconnection/turbulence cascade.
An additional theoretical lower--limit on the distance $R$ to take into account can be derived from the 
photospheric radius $R_{\rm ph}$ \citep{Daigne2002}, below which the emitting region 
is opaque. 
Given the isotropic luminosity of each burst and assuming a typical prompt 
efficiency of $\eta=0.2$ \citep{Nava2014, Beniamini2015}, we checked that the values of $R$ derived using the pair--production argument for each bulk Lorentz factor obtained 
are safely above the associated values of the photospheric radius (in the range $R_{\rm ph} \sim 1.6 \times 10^{12} - 6.8 \times 10^{13}$ cm).

On the other side, $R$ is limited also downward by the variability timescale of the prompt light curve. 
Assuming the standard scenario where the emission is produced by expanding spherical shells, 
the short (order of approximately tens or hundreds of ms, see e.g., \citet{Golkhou2015}) 
variability timescales observed in the prompt emission constrain the possible 
combinations of bulk Lorentz factor $\Gamma$ and distance $R$ from the central 
engine (i.e. $t_{\rm var} = R/2 c \Gamma^2 \sim 0.167 \, R_{14}/2 c \Gamma_{2}^2 $ s). 
If the prompt emission site is moved at larger distances from the central engine, 
for example $R\sim 10^{16}$ cm, it becomes hard to obtain the observed fast variability timescales without invoking extreme bulk Lorentz factors (i.e. $\Gamma \sim 1000$). A somewhat similar way out has been proposed in magnetic dissipation models (e.g., ICMART, \citealt{Zhang2011}), where mini-jets moving at $\Gamma_{\rm mini-jets}$ are used to produce the short observed variability while keeping the emitting region at large radii ($R\sim 10^{15-16}$ cm). Nevertheless, the recent findings on the low-energy breaks \citep{gor2017a,gor2018,Ravasio2018,Ravasio2019} require a small magnetic field ($B^{\prime} \sim$1-10 G, in the leptonic scenario of the synchrotron interpretation) in the emitting region, and the idea of a magnetically dominated jet appears difficult to be reconciled with such low values (for other constraints on GRB prompt emission models based on Poynting-flux-dominated jets, see also \citealt{Beniamini2014}).\\

\begin{table}[h]
    \centering
    \caption{Bulk Lorentz factors $\Gamma$ for each GRB that was co-detected by GBM and LAT and analyzed in this work, applying the compactness argument. Depending on the presence or absence of a cutoff in the high-energy spectrum, these values are measurements or lower--limits, respectively. For each GRB, the table list the name, the method applied to derive $\Gamma$, the corresponding value of bulk Lorentz factor $\Gamma$, the implied value of the distance R from the central engine and the variability timescale, which we assumed to be 100 ms for each GRB.
    For the only two GRB in our sample (GRB~080916C and GRB~110131) with also an independent estimate of $\Gamma$ derived from the afterglow onset in \citealt{Ghirlanda2018}, we reported those $\Gamma$ values for the ISM and wind--like cases and the implied lower--limits on the distance R when combined with the pair--production opacity argument (following Eq.\ref{eq:cutoff_R}). In those two cases, we also report the different variability timescale implied by the corresponding values of R and $\Gamma$, following $t_{\rm var} = R(1+z)/(2c \Gamma^2)$.
    }
    \label{tab:cutoff_R}
    \footnotesize
    \begin{adjustbox}{max width=\columnwidth}
    \begin{tabular}{ccccc}
    \hline
                \multicolumn{1}{c}{\multirow{2}{*}{Name}} &
                \multicolumn{1}{c}{\multirow{2}{*}{Method}} &
                \multicolumn{1}{c}{\multirow{2}{*}{$\Gamma$}} &
                \multicolumn{1}{c}{\multirow{2}{*}{R [cm]} } &
                \multicolumn{1}{c}{\multirow{2}{*}{$t_{\rm var}$ [ms]}} \vspace{0.3cm} \\
                \hline\hline
    \multirow{2}{*}{110721} & \multirow{2}{*}{$\gamma\gamma \xrightarrow{} e^{+}e^{-}$} & \multirow{2}{*}{$>412.7$} & \multirow{2}{*}{$> 3.41\times10^{14}$} & \multirow{2}{*}{100}\\
    \multirow{2}{*}{} & \multirow{2}{*}{} & \multirow{2}{*}{} & \multirow{2}{*}{} & \multirow{2}{*}{(assumed)} \vspace{0.3cm}\\
    \hline
    \vspace{0.1cm} 160625 & $\gamma\gamma \xrightarrow{} e^{+}e^{-}$ & $283.3_{-9.4}^{+8.5}$ & $ 2.00_{-0.13}^{+0.12}\times10^{14}$ & 100 \\
    \hline
    \multirow{3}{*}{} & \multirow{3}{*}{$\gamma\gamma \xrightarrow{} e^{+}e^{-}$} & \multirow{3}{*}{$>745.1$} & \multirow{3}{*}{$> 6.22\times10^{14}$} & \multirow{3}{*}{100 } \vspace{0.1cm}\\         
    \multirow{3}{*}{080916} & \multirow{3}{*}{Aft. onset (ISM)} & \multirow{3}{*}{$1410\pm151$} & \multirow{3}{*}{$> 3.73\times10^{13}$ [Eq.~\ref{eq:cutoff_R}]} & \multirow{3}{*}{$>1.67$} \vspace{0.1cm}\\
    \multirow{3}{*}{} & \multirow{3}{*}{Aft. onset (Wind)} & \multirow{3}{*}{$660\pm 49$} & \multirow{3}{*}{$> 1.10\times10^{15}$ [Eq.~\ref{eq:cutoff_R}]} & \multirow{3}{*}{$>225$} \vspace{0.5cm}\\
    \hline
    170214 & $\gamma\gamma \xrightarrow{} e^{+}e^{-}$ & $333.8_{-17.2}^{+22.0}$ & $ 1.89_{-0.19}^{+0.26}\times10^{14}$ & 100 \vspace{0.1cm}\\
    \hline
    100724 & $\gamma\gamma \xrightarrow{} e^{+}e^{-}$ & $203.9_{-10.7}^{+8.8}$ & $ 0.83_{-0.08}^{+0.07}\times10^{14}$ & 100 \vspace{0.1cm}\\
    \hline
    140206 & $\gamma\gamma \xrightarrow{} e^{+}e^{-}$ & $257.5_{-7.6}^{+9.5}$ & $ 1.33_{-0.08}^{+0.10}\times10^{14}$ & 100 \vspace{0.1cm}\\
    \hline
    131108 & $\gamma\gamma \xrightarrow{} e^{+}e^{-}$ & $>398.1$ & $ >2.80\times10^{14}$ & 100 \vspace{0.1cm}\\
    \hline
     141028 & $\gamma\gamma \xrightarrow{} e^{+}e^{-}$ & $>231.8$ & $ >0.97\times10^{14}$ & 100 \vspace{0.1cm}\\
    \hline
    100116 & $\gamma\gamma \xrightarrow{} e^{+}e^{-}$ & $>278.0$ & $ >1.54\times10^{14}$ & 100 \vspace{0.1cm}\\
    \hline
    160821 & $\gamma\gamma \xrightarrow{} e^{+}e^{-}$ & $307.5_{-7.9}^{+7.6}$ & $ 1.89_{-0.10}^{+0.10}\times10^{14}$ & 100 \vspace{0.1cm}\\
    \hline
    130504 & $\gamma\gamma \xrightarrow{} e^{+}e^{-}$ & $>136.2$ & $ >0.37\times10^{14}$ & 100 \vspace{0.1cm}\\
    \hline
    110328 & $\gamma\gamma \xrightarrow{} e^{+}e^{-}$ & $>201.8$ & $ >0.81\times10^{14}$ & 100 \vspace{0.1cm}\\
    \hline
    160509 & $\gamma\gamma \xrightarrow{} e^{+}e^{-}$ & $228.3_{-6.3}^{+6.1}$ & $ 1.44_{-0.08}^{+0.08}\times10^{14}$ & 100 \vspace{0.1cm}\\
    \hline
    180720 & $\gamma\gamma \xrightarrow{} e^{+}e^{-}$ & $129.8_{-2.8}^{+2.4}$ & $ 0.61_{-0.03}^{+0.02}\times10^{14}$ & 100 \vspace{0.1cm}\\
    \hline
    151006 & $\gamma\gamma \xrightarrow{} e^{+}e^{-}$ & $>229.5$ & $ >1.05\times10^{14}$ & 100 \vspace{0.1cm}\\
    \hline
    160905 & $\gamma\gamma \xrightarrow{} e^{+}e^{-}$ & $176.6_{-7.3}^{+10.5}$ & $ 0.62_{-0.05}^{+0.08}\times10^{14}$ & 100 \vspace{0.1cm}\\
    \hline
    150902 & $\gamma\gamma \xrightarrow{} e^{+}e^{-}$ & $>254.6$ & $ >1.30\times10^{14}$ & 100 \vspace{0.1cm}\\
    \hline
    090328 & $\gamma\gamma \xrightarrow{} e^{+}e^{-}$ & $143.5_{-5.2}^{+5.2}$ & $ 0.71_{-0.05}^{+0.05}\times10^{14}$ & 100 \vspace{0.1cm}\\
    \hline
    100826 & $\gamma\gamma \xrightarrow{} e^{+}e^{-}$ & $>282.0$ & $ >1.6\times10^{14}$ & 100 \vspace{0.1cm}\\
    \hline
    \multirow{3}{*}{} & \multirow{3}{*}{$\gamma\gamma \xrightarrow{} e^{+}e^{-}$} & \multirow{3}{*}{$>495.3$} & \multirow{3}{*}{$> 3.84\times10^{14}$} & \multirow{3}{*}{100 } \vspace{0.1cm}\\
                            
   \multirow{3}{*}{110731}  & \multirow{3}{*}{Aft. onset (ISM)} & \multirow{3}{*}{$971.2\pm12.1$} & \multirow{3}{*}{$> 1.60\times10^{13}$ [Eq.~\ref{eq:cutoff_R}]} & \multirow{3}{*}{$>1.08$} \vspace{0.1cm}\\
    
    \multirow{3}{*}{} & \multirow{3}{*}{Aft. onset (Wind)} & \multirow{3}{*}{$331.5\pm 8.3$} & \multirow{3}{*}{$> 2.57\times10^{15}$ [Eq.~\ref{eq:cutoff_R}]} & \multirow{3}{*}{$>1494$} \vspace{0.5cm}\\
    \hline
    160910 & $\gamma\gamma \xrightarrow{} e^{+}e^{-}$ & $>289.6$ & $ >1.68\times10^{14}$ & 100 \vspace{0.1cm}\\
    \hline
    150202 & $\gamma\gamma \xrightarrow{} e^{+}e^{-}$ & $>171.4$ & $ >0.59\times10^{14}$ & 100\\
    \hline
    \hline
    \end{tabular}
    \end{adjustbox}
\end{table}

\begin{figure}[h]
    \centering
    \includegraphics[width=\columnwidth]{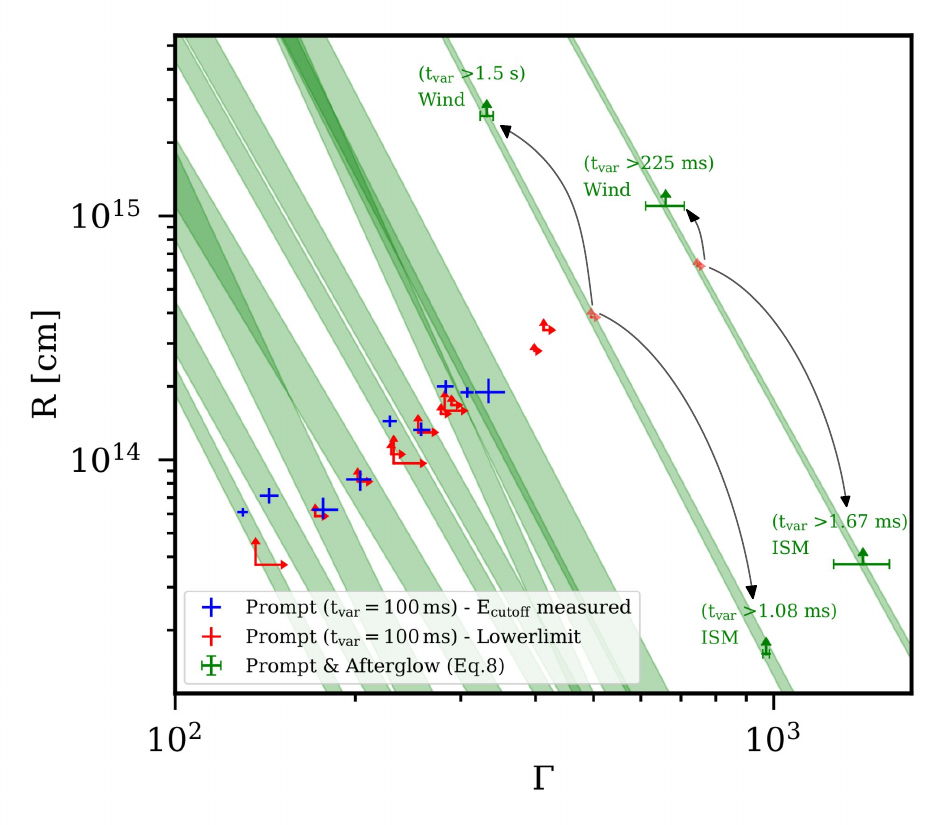}
\vskip -0.3 cm
    \caption{Constraints on the bulk Lorentz factor $\Gamma$ and distance $R$ of the emitting region from the central engine, for the 22 bright GRBs analyzed in this work. The measured values (blue crosses) or lower limits (red arrows) on $\Gamma$, and consequently on $R$ (assuming $t_{\rm var} = 100$~ms), are derived according to the presence or absence of a high-energy cutoff in the spectrum, respectively. 
    For each GRB with a measured value of $\Gamma$, the green stripes represent the region of the parameter space allowed by the compactness argument for any given $\Gamma$, using only the prompt spectral properties and following Eq.~\ref{eq:cutoff_R}.
    The combination of the information from the prompt and the afterglow, when available, allows us to break the degeneracy and better constrain the distance $R$.
    The four black arrows represent the cases of GRB 080916 and GRB 110731, where an independent estimate of $\Gamma$ from the afterglow onset was taken from \citealt{Ghirlanda2018}, allowing to put constraints on the distance $R$.
    Corresponding lower limits on variability timescales are provided for each case.
    Each error bar represents the 1-$\sigma$ error on measurements.
    }
    \label{fig:R_G}
\end{figure}

Figure~\ref{fig:R_G} shows a summary of the results obtained in this work on the bulk Lorentz factor $\Gamma$ and the distance $R$ of the emitting region from the central engine (as reported in Table~\ref{tab:cutoff_R}). Depending on whether a cutoff was present or absent in the prompt emission spectrum, we derived measurements (blue points) or lower limits (red arrows) on $\Gamma$, respectively, and, assuming $t_{\rm var} = 100$~ms, on the corresponding value of $R$.
Following Eq.~\ref{eq:cutoff_R} and taking into account the distribution of the spectral parameters, we obtained the range of the parameter space allowed by the compactness argument for each GRB, based only on the study of its prompt spectral properties. The resulting ranges are represented by the green stripes, where, for any given $\Gamma$, the 68\% of the allowed distribution of $R$ is shown (only for measured values, for graphic clarity). The slope of each stripe varies for each GRB, based on its specific $\beta_e$ value (see Eq.~\ref{eq:cutoff_R}). If, as in the cases of GRB 080916 and GRB 110731, an independent estimate of $\Gamma$ is available (in this case taken from \citealt{Ghirlanda2018}), this allows to break the degeneracy and better constrain the values of $R$ on the stripe.  
Indeed, using the values of the bulk Lorentz factors for the ISM and wind-like medium cases from the afterglow, the lower limits of GRB 080916 and GRB 110731 (represented by transparent red arrows) have been transformed into the corresponding four green arrows. 
as in these two cases the absence of a cutoff in the prompt spectrum allowed to set only a lower limit on $R$.
The corresponding lower limits on the variability timescales are reported for each of the four cases.\\

In our analysis, this method for the estimate of the distance $R$ has been applied to 
only two GRBs of our sample which satisfied the criterion of having $\Gamma$ estimated in an independent way. 
However, Eq.~\ref{eq:cutoff_R} is general and can be applied to any other GRB which has the bulk Lorentz factor $\Gamma$ inferred in a independent way and whose spectral parameters of the prompt emission are known. 
Indeed, this method has been recently applied in \cite{Mei2022} to GRB 220101A, a bright burst ($E_{\rm iso} \sim 3 \times 10^{54}$ erg) detected by \fe/GBM and \fe/LAT. 
The joint multiwavelength analysis of the prompt emission spectrum allowed to characterize the full spectral shape from X--rays to GeV energies, revealing the detection of a spectral cutoff at $E_{\rm cutoff} =80^{+44}_{-14}$ MeV (similar results found also in \citealt{Scotton2023}). 
Combining the information gathered from the prompt emission spectrum and the afterglow light curve, the authors derived stringent constraints on the GRB prompt emission site, finding $\Gamma \sim 900$ and $R \sim 4.5 \times 10^{13}$ cm (see their Figure 4). 
We note that this GRB did not belong to the \fe/LAT catalog at the time of the selection of the sample for this work. 
However, while it would have satisfied the criterion set on 
the fluence detected in GBM (its fluence being $F = 6.03 \pm 0.02 \times 10^{-5} \,\,\rm erg/cm^2$), its significance in LLE data is 
only $\sigma = 14$, so it would not be part of our selected sample of bright 
LLE-detected bursts. 
Nonetheless, this burst clearly showed evidence of a cutoff at high--energies, 
implying that there might be other GRBs candidates with a cutoff in their prompt spectrum outside our S/N-selected sample.

\section{Conclusions}\label{sect:conclusions}

In this work, we present the spectral analysis of the 22 brightest GRBs simultaneously detected by both instruments on board Fermi, GBM, and LAT, covering a wide energy range, from 10 keV to 10 GeV.
The joint analysis of GBM and LAT data for the 22 time-integrated prompt emission GRB spectra revealed that in nine bursts the spectral data significantly deviate from the extrapolation of the high--energy power law, requiring the presence of an exponential cutoff at high energies, between $\sim$14 and 298 MeV.
These results are consistent with the previous detections of exponential cutoffs in GRB prompt spectra reported in a few cases in the literature \citep[e.g.,][]{Vianello2018,Tang2015}. 
In the remaining 13 out of 22 GRBs of the sample, there is no significant evidence of a spectral cutoff in the data. 
Since the poor signal in the highest part of the spectrum might hamper the evidence of a cutoff, for each of the 13 GRBs we derived a lower limit on the cutoff position.

Interpreting the exponential cutoff observed in nine out of 22 GRBs as a sign of the pair-production opacity, we derived the estimate (and the lower limit for the remaining 13 GRBs not showing the cutoff) of the jet bulk Lorentz factor $\Gamma$ during the prompt phase  (see Fig.~\ref{fig:Gamma0}).
The distribution of the estimates (lower limits) of $\Gamma$ inferred from the pair-production argument spans the range of values 130--333 (136--745), which are compatible with the typical values assumed in the GRB standard model and are consistent with those derived from the same opacity argument reported in the literature.

By comparing the $\Gamma$ values inferred from the opacity argument reported in this work with the ones inferred from the afterglow onset time reported in \citealt{Ghirlanda2018}, we found that our results are consistent with the distribution of $\Gamma$ derived for the homogeneous medium case (see Fig.~\ref{fig:cutoff_gamma0distrib}). 
In the wind case, instead, the distribution of bulk Lorentz factors extends toward lower values of $\Gamma$, which would imply lower values of the cutoff energy, and consequently a fainter flux in the LAT range. Therefore, the absence of such lower values in the $\Gamma$ distribution of LAT-detected bursts could be due to a selection effect hampering the consistency with the bulk Lorentz factors derived in the wind case. 
The distribution of $\Gamma$ may also extend toward higher values, as is indeed suggested by the lower limits on $\Gamma$ values, but their observation is hampered by the limited energy range of the current instruments.\\

Combining all the spectral fits performed in this analysis, we derived the distribution of the slope $\beta$ of the high-energy power law, with median value $\langle \beta \rangle$ = -2.39 ($\sigma =0.56$). By comparing it with the distribution of $\beta$ from the \fe/GBM catalog, we found steeper $\beta$ values with respect to the ones reported in the catalog. This could be an indication that the true slope of the GRBs' prompt emission spectrum is softer than what is usually found, and it is likely due to the addition of the LLE and LAT data in our analysis, which allowed us to better constrain the high--energy power--law slope of the spectrum.

From the slopes $\beta$ of the power law modeling the high--energy spectrum of the 22 GRBs analyzed, we derived the corresponding slopes $p$ of the underlying nonthermal distribution of accelerated particles, assuming synchrotron emission in the fast-cooling regime. 
The distribution of $p$ is quite broad and does not cluster around a universal value.
We found that the distribution of $p$ has a median value of $p=2.79$ with a tail reaching values of $p \sim$ 5--7. 
Steep values of $p$, coupled also with the large values of $\gamma_{\rm min}$ found from independent synchrotron modeling of other bursts in the literature (e.g., \citealt{Ronchi2020,Ryde2022}), require an efficient acceleration mechanism (mildly relativistic shocks or magnetic reconnection) giving rise to a steep energy distribution of accelerated particles.
Given the theoretical uncertainties on the energy distribution of accelerated particles in mildly relativistic shocks, these results provide useful observational benchmarks for the development of the theory of particle acceleration applied to the prompt emission case.\\

Our results show how crucial the inclusion is, when possible, of the LLE (30--100 MeV) and LAT data (>100 MeV) to the study of a GRB prompt spectrum detected by the GBM (typically sampled in the energy range 10 keV--10 MeV).  
A broadband study of the prompt emission spectrum can reveal significant deviations from the standard spectral shape often found when analyzing the spectra in a narrower energy range and with simpler functions (see the solid blue line in Fig.~\ref{fig:gbm_lat_coverage}). 
As we show in this work, once the crucial information from the high-energy prompt emission data is obtained, one can exploit their full potential by combining them with the information derived from the onset of the afterglow component, which allows the location of the emitting region to be derived without using the additional assumption on the variability timescale (see also \citealt{Gupta2008}).
In the only two GRBs of our sample satisfying the criteria, we used the independent 
estimates of $\Gamma$ derived from the afterglow onset reported in 
\citealt{Ghirlanda2018}. 
Depending on the assumptions on the homogeneous or wind medium cases, we found values of 
$R > 10^{13}-10^{15}$ cm, which are compatible with the location of the prompt emission 
region predicted by the GRB standard model, although they may suggest larger distances, as expected in magnetic dissipation models \citep{Lyutikov2003, Zhang2011} and also found in \citealt{Zhang2009} and \citealt{Beniamini2018}, for example. 
 
This method provides a powerful tool to restrict the range of distance $R$ of the prompt emitting region from the central engine, a key parameter in the GRB standard model still suffering from large uncertainties.
Given the encouraging results presented in this work (and also applied in \citet{Mei2022}), we plan to apply this method to other GRBs that satisfy the criteria, namely whose prompt emission has been detected by LAT and with a measured afterglow onset time.
This would significantly help in constraining the allowed region of the parameters' space for prompt emission and distinguishing between different dissipation models proposed in the literature, which invoke different emitting region distances and attainable bulk Lorentz factors. 
The investigation of these interesting issues will be the subject of a forthcoming publication.

\begin{acknowledgements}
We would like to thank the anonymous referee for carefully reading our manuscript and for providing valuable comments, which improved our manuscript.
M.\,E.\,R. is thankful to the Observatory of Brera-Merate for the kind hospitality during the development of this project, and to Dr. Lara Nava for the insightful discussions that helped shape this project. This work has been partially supported by the research programme Athena with project number 184.034.002, which is financed by the Dutch Research Council (NWO).
This research has made use of data obtained through the High Energy Astrophysics Science Archive Research Center Online Service, provided by the NASA/Goddard Space Flight Center, and specifically, this work made use of public \fe/GBM and \fe/LAT data. 
\end{acknowledgements}

\bibliographystyle{aa} 
\bibliography{references} 

\begin{appendix}
\section{Time-integrated spectrum of GRBs showing a cutoff}
\begin{figure*}[h]
\centering
\includegraphics[scale=0.16]{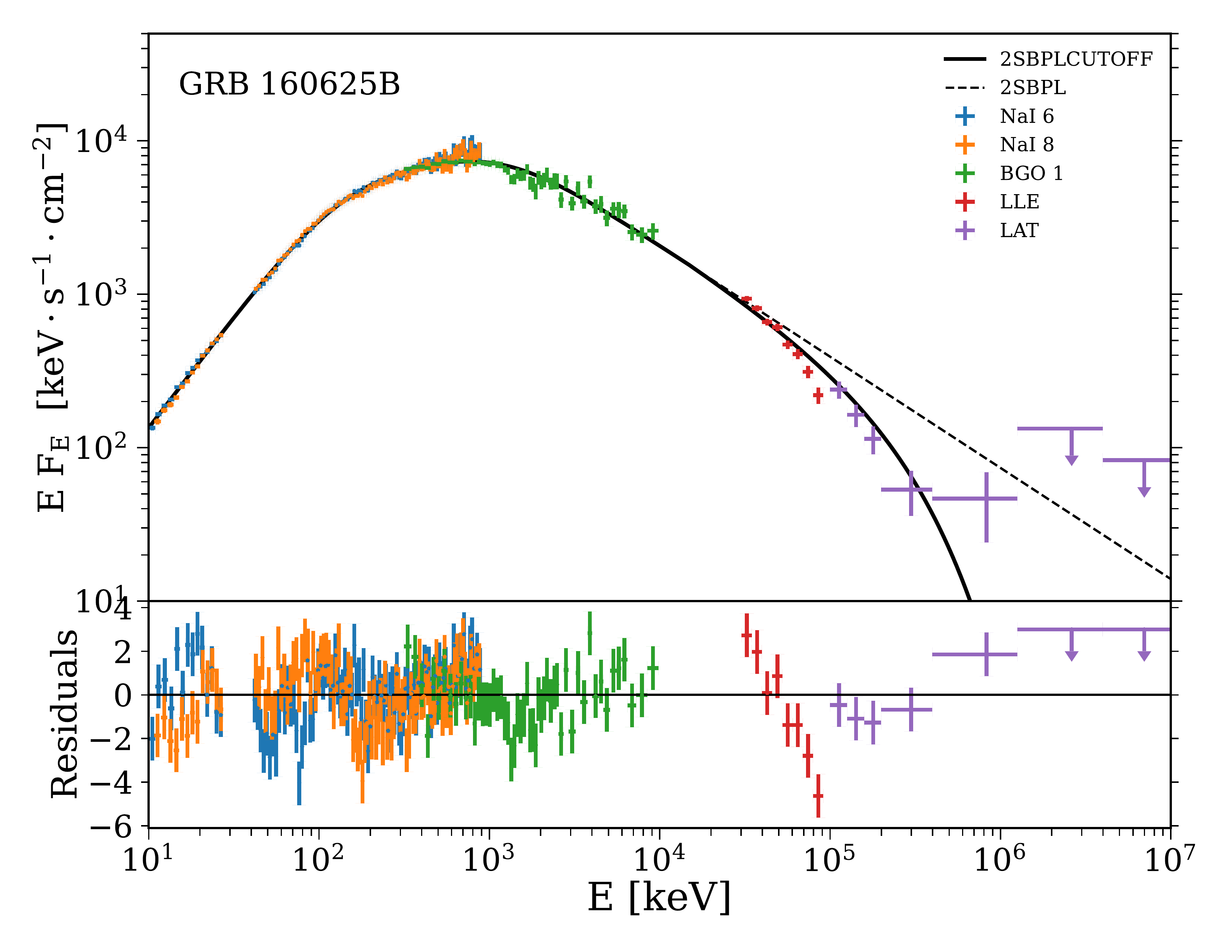} 
\includegraphics[scale=0.16]{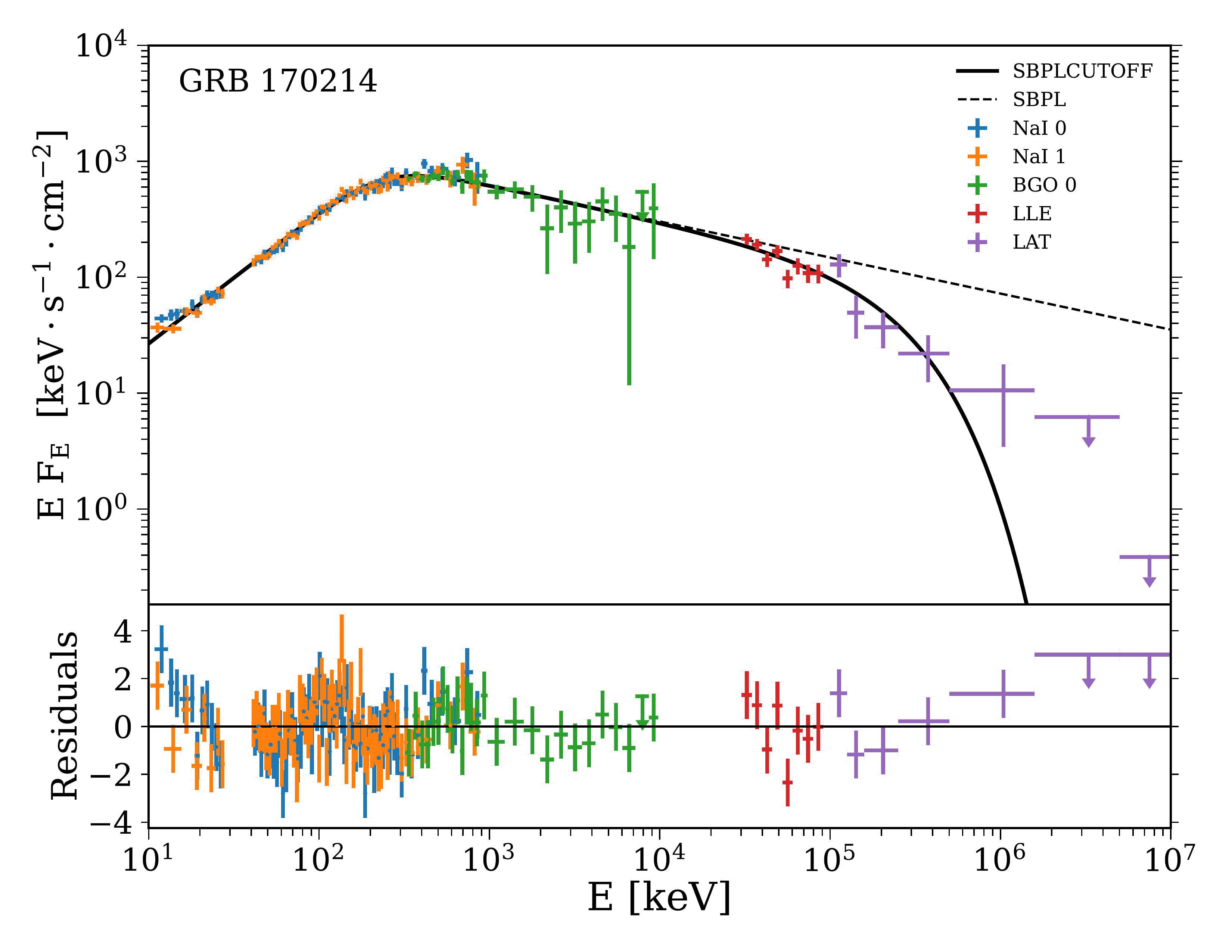}  
\includegraphics[scale=0.16]{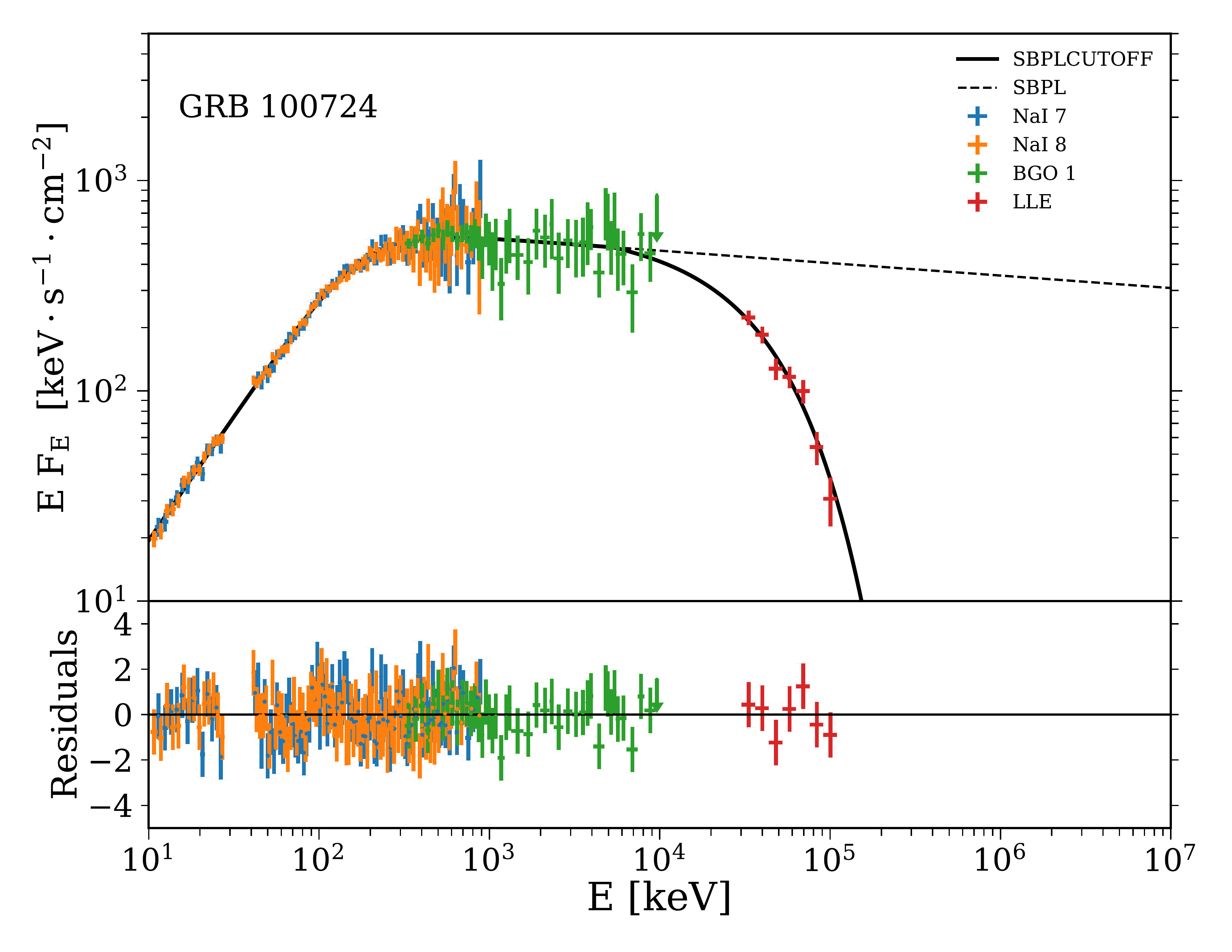} 
\includegraphics[scale=0.16]{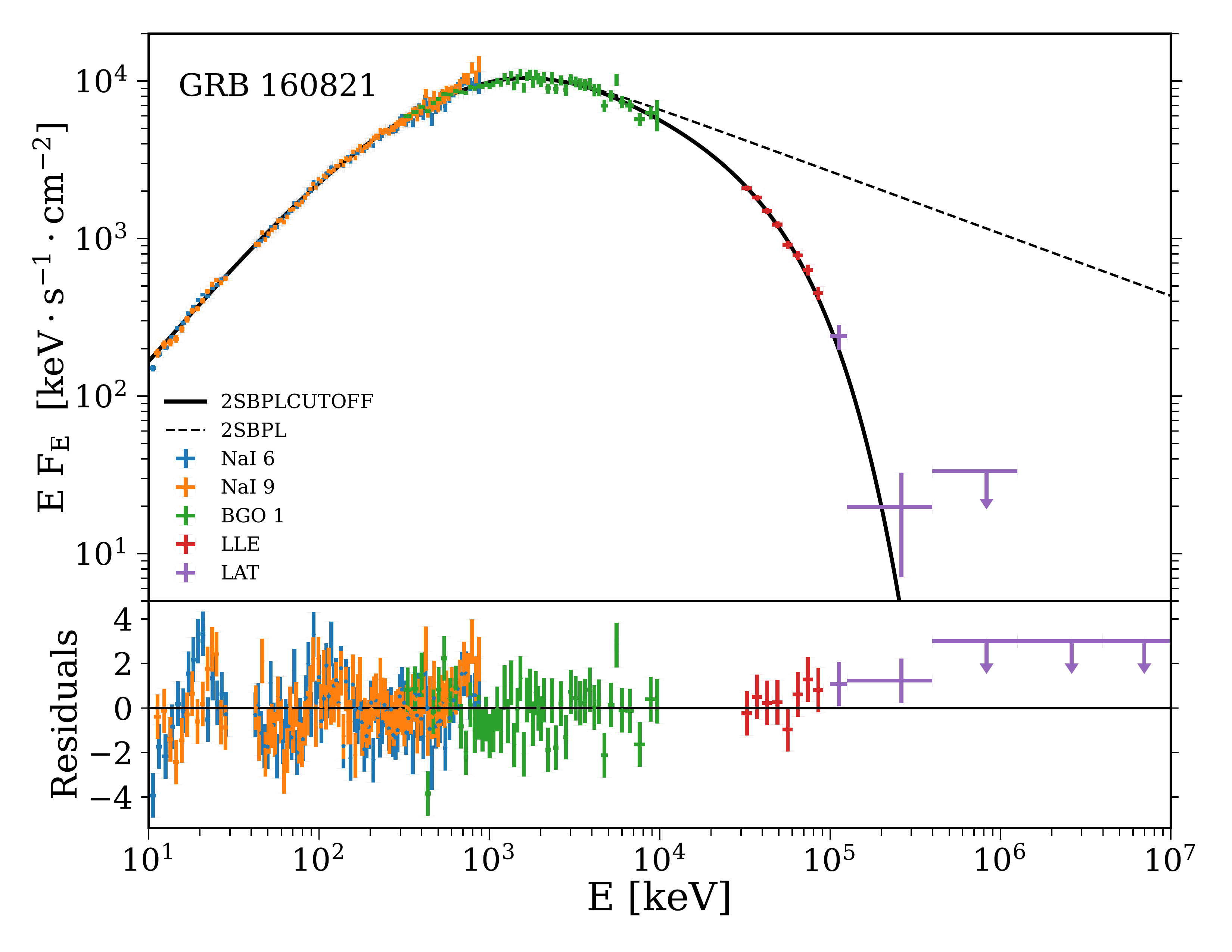}
\includegraphics[scale=0.16]{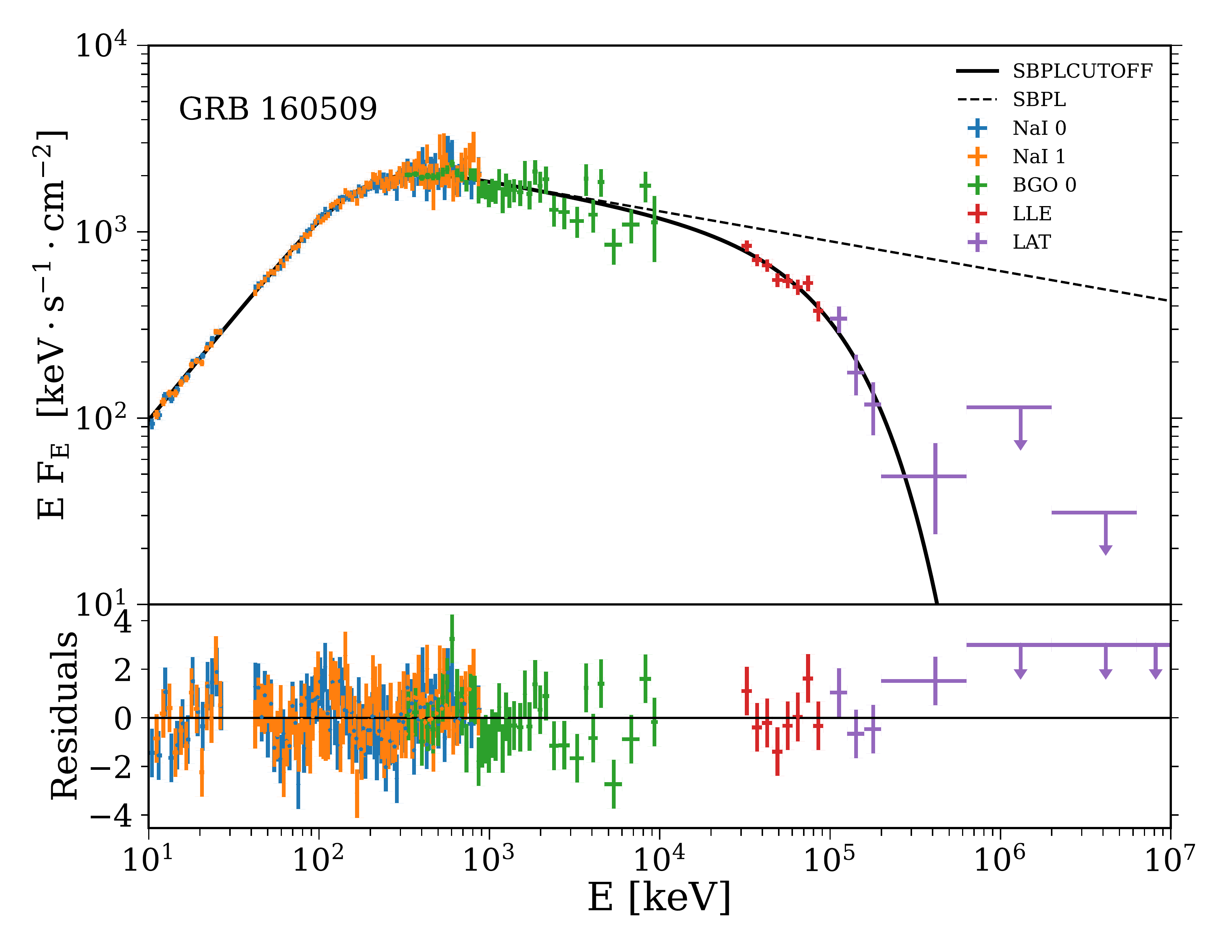} 
\includegraphics[scale=0.16]{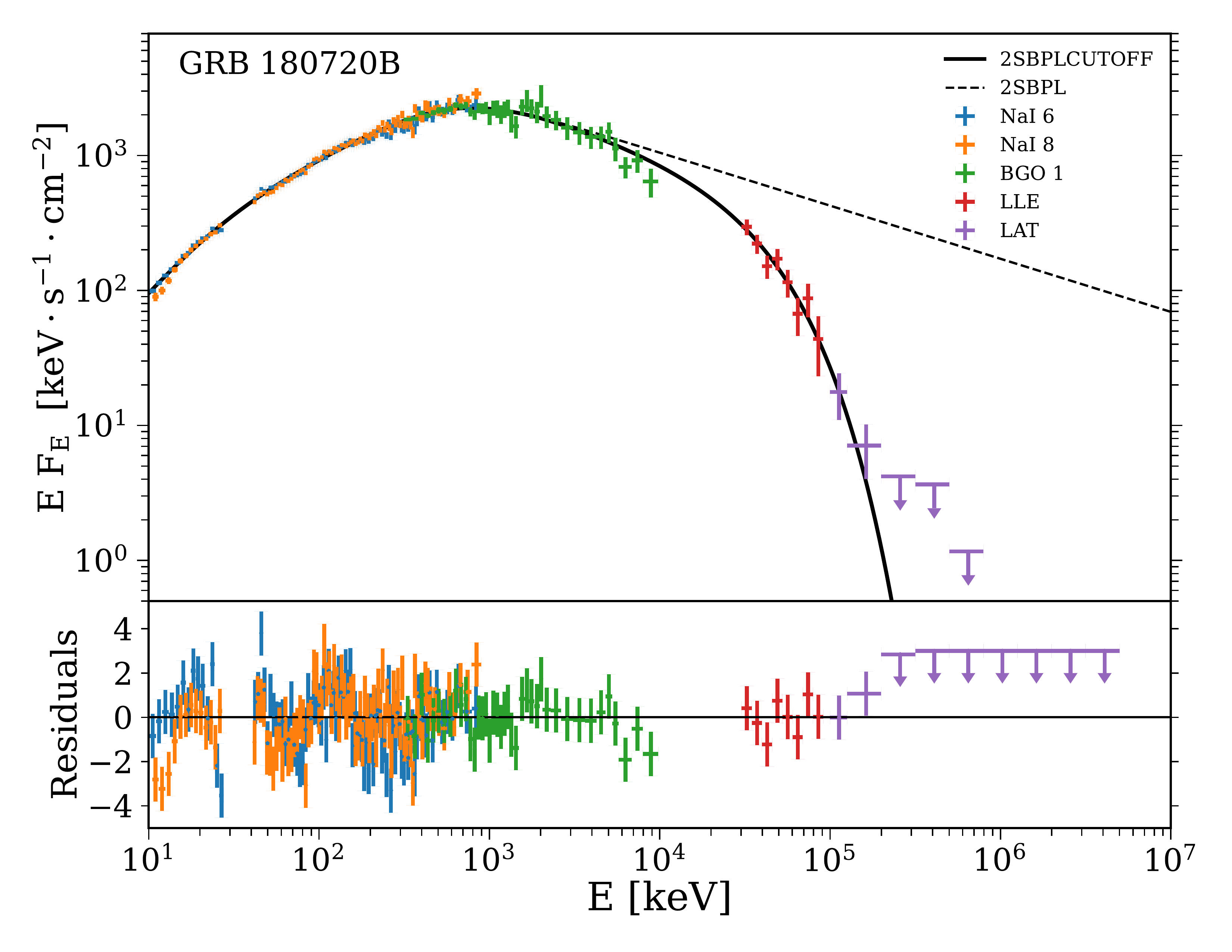} 
\caption{Time-integrated spectra of the GRBs showing the presence of an exponential cutoff at high energies, i.e., those that are best fitted by either the SBPLCUTOFF or 2SBPLCUTOFF function (solid line), excluding GRB 140206, whose spectrum has already been reported in Fig.~\ref{fig:cutoff_140206}. The best-fit values of the corresponding models are reported in Table~\ref{tab:cutoff_bestfitparams}. For each plot, data correspond to the signal detected from 10 keV to 10 GeV by the different instruments on board Fermi (color-coded as shown in the legend). Data have been rebinned for graphical purposes.  }
\label{Appendix_fig}
\end{figure*}
\setcounter{figure}{0}
\begin{figure}[h]
\centering
\includegraphics[scale=0.16]{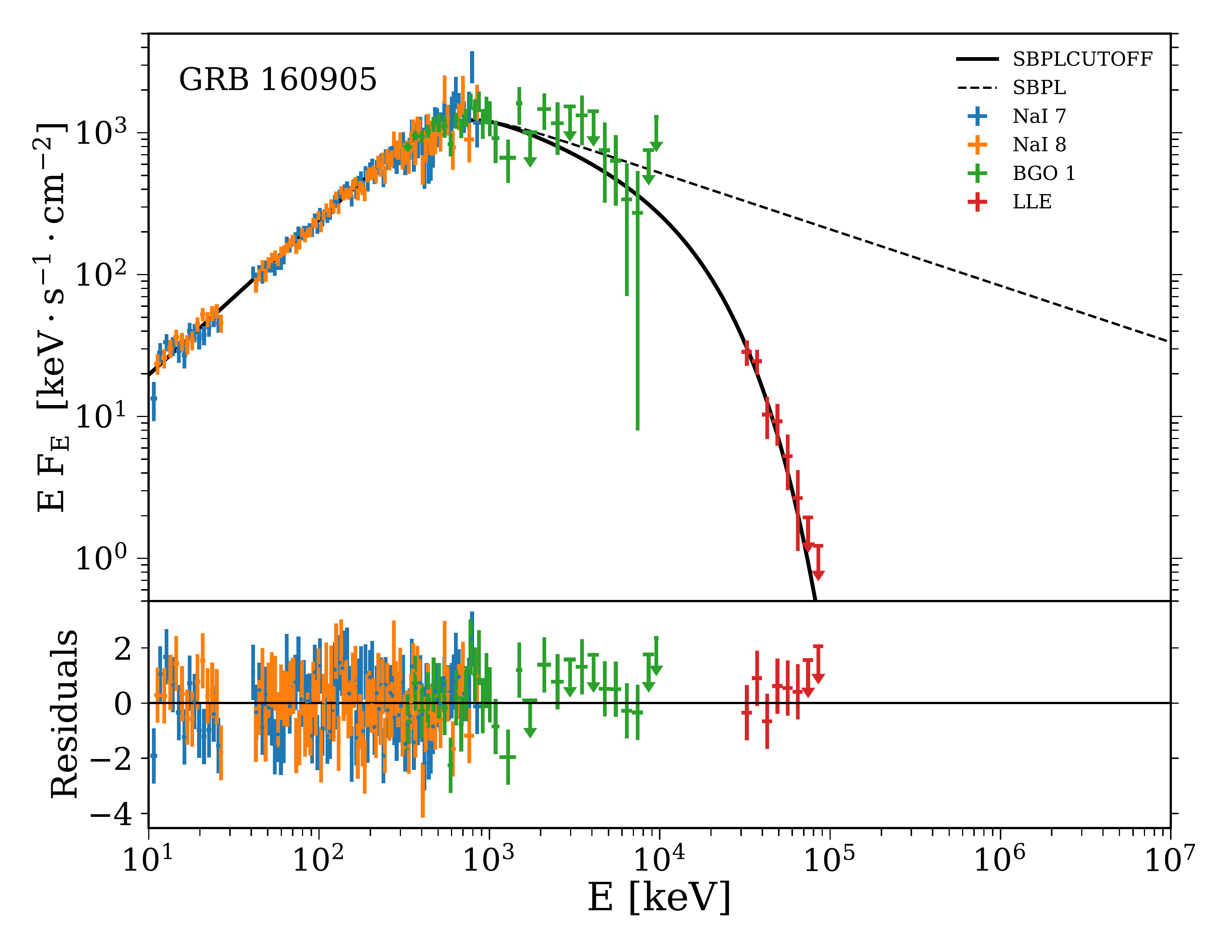} 
\includegraphics[scale=0.16]{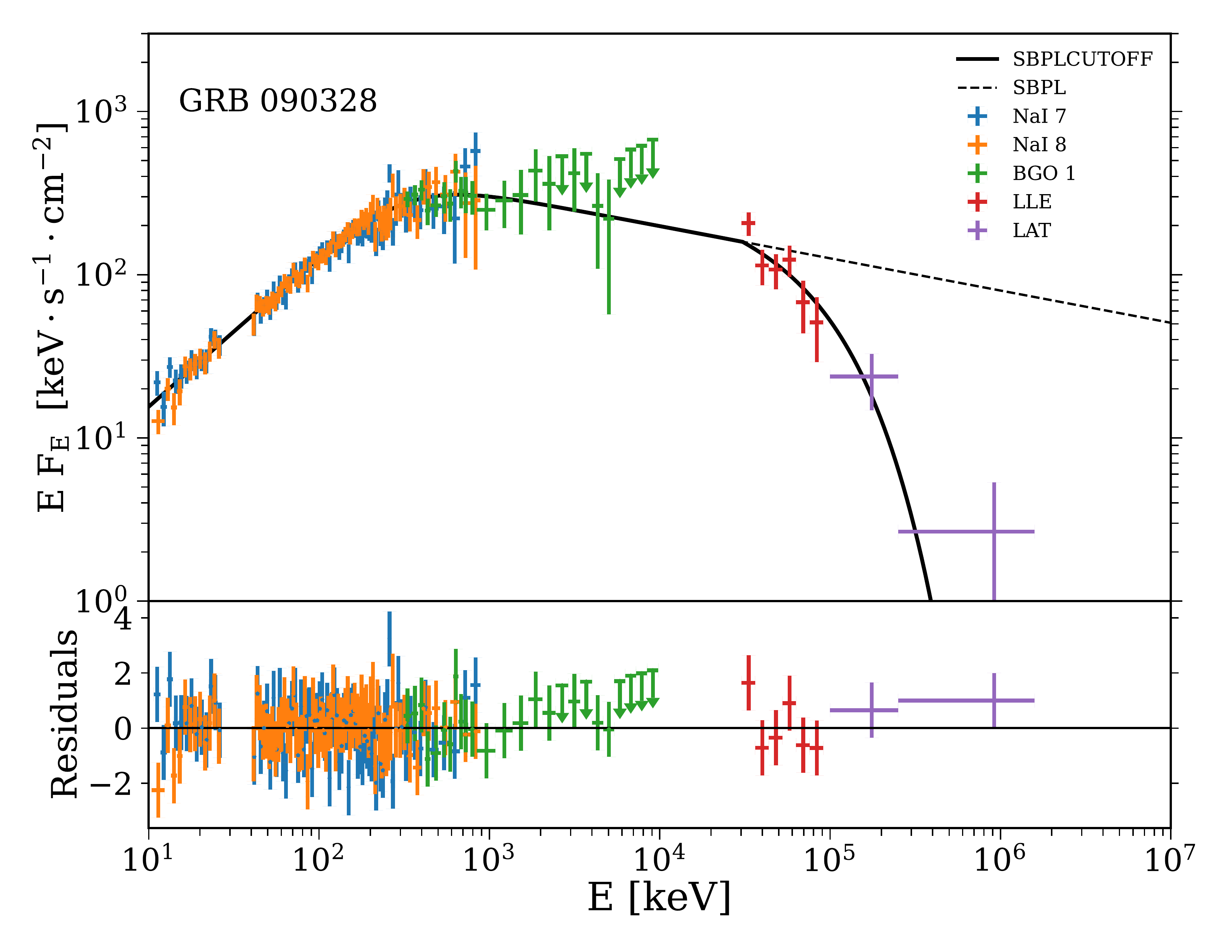} 
\caption{(Continued)}
\end{figure}

\end{appendix}


\end{document}